%% file: main.tex
\AddToHook{package/cleveref/before}{\RequirePackage[hidelinks,hyperfootnotes=false]{hyperref}}
\documentclass[conference]{IEEEtran}
\input{preamble}
\title{Structure and Complexity of\\ Perfect Functional Representations}

\author{
\IEEEauthorblockN{Serhat Emre Coban, Yanina Y. Shkel, and Emre Telatar}
}
\begin{document}
\maketitle 
\thispagestyle{plain}

\authorlistandfunding{Serhat Emre Coban, Yanina Y. Shkel, and Emre Telatar are with the School of Computer \& Communication Sciences, École Polytechnique Fédérale de Lausanne (EPFL), Switzerland. Email: serhat.coban@epfl.ch, yanina.shkel@epfl.ch, emre.telatar@epfl.ch. This work is funded by the Swiss NSF grant number 211337. This work was presented in part at the 2026 IEEE International Symposium on Information Theory (ISIT)~\cite{CobanShkelTelatar2026}.}

\begin{abstract}
For given random variables $(X,Y)$, functional representation lemma states that there exists a random variable $Z$ and a deterministic function $g$ such that $Y = g(X,Z)$, and $X$ is independent of $Z$. In other words, $Z$ captures all the information about $Y$ not contained in $X$. We refer to $Z$ as a `perfect functional representation' if it satisfies $I(Y;Z) = H(Y|X)$. We show that the existence of perfect representation depends solely on the conditional distribution $p_{Y|X}$ and define the corresponding conditional distributions as perfectly representable. We provide a necessary and sufficient condition for $p_{Y|X}$ to be perfectly representable. Moreover, we cast the problem of checking perfect representability as a linear program, give an algorithm whose complexity is polynomial in $|\mathcal Y|$ and exponential in $|\mathcal X|$, and prove that deciding perfect representability is NP-complete for rational input matrices encoded in binary.
\end{abstract}

\input{final_manuscript}

\section{Computational Complexity}\label{computational complexity chapter}
\input{computational_complexity_v2.tex}

\section{Discussion on Some Information Measures}\label{relations to infor measures section}
\input{common_info}
\section{Concluding Remarks}\label{concluding remarks section}
\balance
\input{concluding_remarks.tex}

\clearpage

\newpage

\bibliographystyle{IEEEtran}
\bibliography{IEEEabrv,References}

\newpage
\appendices
\section{Proofs for Section~\ref{sec:system:model}}\label{appendix A}
\input{appendixA.tex}

\section{Proofs for Section~\ref{sec:main:results}}\label{appendix B}
\input{appendixB.tex}

\section{Proofs for Section~\ref{algorithmic results chapter}}\label{appendix C}
\input{appendixC.tex}

\section{Proofs for Section~\ref{computational complexity chapter}}\label{appendix D}
\input{appendixD.tex}

\section{Proofs for Section~\ref{relations to infor measures section}}\label{appendix E}
\input{appendixE.tex}

\end{document}

%% file: preamble.tex
\IEEEoverridecommandlockouts
\usepackage{cite}
\usepackage[cmex10]{amsmath}
\usepackage{amsmath,amssymb,amsfonts}
\usepackage{graphicx}
\usepackage{textcomp}
\usepackage{xcolor}
\def\BibTeX{{\rm B\kern-.05em{\sc i\kern-.025em b}\kern-.08em
    T\kern-.1667em\lower.7ex\hbox{E}\kern-.125emX}}
\usepackage[utf8]{inputenc} 
\usepackage[T1]{fontenc}
\usepackage{url} 
\usepackage{amsthm}
\usepackage{mleftright}       %
\mleftright                   %

\usepackage{mleftright}       %
\mleftright                   %
\usepackage{graphicx}
\usepackage{booktabs} 
\usepackage[symbol]{footmisc}
\usepackage{dsfont}
\usepackage{xcolor}
\usepackage{graphicx} 
\usepackage{enumerate}
\usepackage{color}
\usepackage{url}
\usepackage{balance} 
\usepackage{mathrsfs}
\usepackage{epsfig}
\usepackage{verbatim}
\usepackage{setspace}
\usepackage{cite}
\usepackage{bbm}
\usepackage{array}
\usepackage{etoolbox}
\usepackage{algorithm}
\usepackage{multicol}
\usepackage{lipsum}
\usepackage{cleveref}
\crefname{step}{Step}{Steps}
\usepackage{etoolbox}
\usepackage{algpseudocode}

\renewcommand{\vec}[1]{\mathbf{#1}}

\renewcommand{\thefootnote}{\fnsymbol{footnote}}
\newcommand{\authorlistandfunding}[1]{%
\begingroup%
\let\thefootnote\relax%
\footnotetext{#1}%
\endgroup%
}
\renewcommand{\thefootnote}{\fnsymbol{footnote}}

\makeatletter
\patchcmd{\@IEEEeqnarray}{\relax}{\relax\intertext@}{}{}
\makeatother

\newtheorem{theorem}{Theorem}
\newtheorem{lemma}{Lemma}

\newtheorem{corollary}{Corollary}
\newtheorem{definition}{Definition}

\newtheorem{remark}{Remark}
\newtheorem{example}{Example}

%% file: final_manuscript.tex
\section{Introduction}\label{sec:introduction}
For given random variables $(X,Y)$, functional representation lemma~\cite{elgamal2011network} states that there exists a random variable $Z$ and a deterministic function $g$ such that $Y = g(X,Z)$, and $X$ is independent of $Z$. In other words, the triple $(X,Y,Z)$ satisfies
\begin{align} 
I(X;Z) &= 0  \quad \mbox{and} \quad H(Y|X,Z) = 0.\label{functional representation restrictions} 
\end{align}
A random variable $Z$ that satisfies~(\ref{functional representation restrictions}) is not unique. The problem of finding a $Z$ with minimum entropy has been studied extensively in~\cite{cicalese,compton1,compton2,yana,yana2,cheuk,rossi,mir,sy,sy2,eci2,eci3} and references therein. The problem of finding $Z$ maximizing the mutual information with $Y$ has been a key technical ingredient in information theory problems such as channel synthesis~\cite{SFRL,PML,CCoC,sriramu2024optimalredundancyexactchannel}, and many others~\cite{SFRL,PML}. Finally, there are many interesting connections between functional representation and information-theoretic privacy~\cite{yana, yana2, Zamani, Zamani2, Zamani3, Zamani4, Zamani5, Zamani6, Liu2024UniversalExactCompression, Liu2024OneShotInformationHiding,yana_zurich_seminar}.

As mentioned above, finding a $Z$ that is as informative as possible about $Y$ is of particular interest that is to maximize $I(Y;Z)$ under the constraints in~(\ref{functional representation restrictions}). Observe that by the chain rule, 
\begin{align}
I(Y;Z) 
= H(Y|X) - I(X;Z|Y)- H(Y|X,Z) + I(X;Z), 
\end{align}
and by~(\ref{functional representation restrictions}) we have
\begin{align}
I(Y;Z) = H(Y|X) - I(X;Z|Y) \leq H(Y|X). \label{trivial upper bound}
\end{align}
Hence, the maximum mutual information is trivially upper-bounded by $H(Y|X)$, with equality whenever $I(X;Z|Y) = 0$. This happens if and only if $X-Y-Z$, i.e., the triple $(X,Y,Z)$ forms a Markov chain.  

It is shown in~\cite{SFRL,CCoC} that it is possible to
approach the upper bound in~(\ref{trivial upper bound}) within
an additive logarithmic term. That is, there exists a $Z$ such that
\begin{align}
H(Y|X) - \log(I(X;Y) + 1) - 4 \leq I(Y;Z). \label{SFRL}
\end{align}
From this, we can easily see that the trivial upper bound in~(\ref{trivial upper bound}) is achieved asymptotically for product distributions. That is, applying the bound in~(\ref{SFRL}) to $(X^n,Y^n)$ gives a functional representation $Z_n$ for which $I(X^n;Z_n|Y^n)$ is $O(\log n)$ (and thus goes to zero if we normalize by $n$). This problem is further studied in the context of channel synthesis in~\cite{sriramu2024optimalredundancyexactchannel} where it is shown that the redundancy term $I(X^n;Z_n|Y^n)$ has leading term $\frac{1}{2}\log n$ for $(X^n,Y^n)$ product pairs with a full-support finite-alphabet non-singular conditional distribution $p_{Y|X}$. Note that we will define singularity in Definition~\ref{def: perfect channels}. Moreover, for product pairs with a singular conditional distribution, $p_{Y|X}$, the redundancy term is shown to be sub-logarithmic.

The goal of this paper is to characterize random variables for which there exists a functional representation $Z$ such that the trivial upper bound in~(\ref{trivial upper bound}) is achieved exactly. We call such pairs $(X,Y)$ perfectly representable and $Z$ a perfect functional representation of the pair $(X,Y)$. This problem is surprisingly ubiquitous. As it is mentioned in~\cite{yana}, there exist a perfect functional representation in some special cases: for example, if $X$ is a function of $Y$ or $Y$ is a function of $X$. In~\cite{SFRL} it is shown that the term $I(X;Z|Y)$ is minimized by the Poisson functional representation whenever the support of $Y$ is two. This problem was further studied in~\cite{Zamani} where it is shown that if Wyner's common information for $(X,Y)$ is the same as the mutual information, then the pair is perfectly representable.
Moreover, it could be observed from \cite{Korner1973,Alon^2} that perfect representability is equivalent to $H_K(G_{YX},Y)=H(Y|X)$ where $H_K$ denotes K\"orner graph entropy and the graph $G_{YX}$ is defined as in \cite{LiElGamal2018EGW}. Likewise, perfect representability is equivalent to perfect privacy in the privacy funnel, i.e., $G_{\mathrm{PF}}(H(Y|X),X\!\to\!Y)=0$ (see\cite{privacy_funnel}). Finally, this problem could be viewed as a special case of the channel synthesis problem in~\cite{sriramu2024optimalredundancyexactchannel}. That is, we characterize exactly when the redundancy term in channel synthesis is zero, and not just vanishing asymptotically.

Our main result gives a necessary and sufficient condition for perfect representability in terms of elementary operations on singular conditional probability matrices. We then cast perfect representability as a linear program. By solving its dual through a separation oracle, we obtain a $2^{|\mathcal X|}\operatorname{poly}(|\mathcal Y|,L)$-time algorithm, where $L$ denotes the bit complexity of the entries of $p_{Y|X}$. We further prove that deciding perfect representability is an NP-complete problem when the input is the rational matrix $p_{Y|X}$, with input size measured by its binary encoding length; our proof uses a polynomial-time reduction from 3-colorability on simple graphs and a polynomial-size certificate for perfect representability.

The remainder of this paper is structured as follows. In Section~\ref{sec:system:model}, we define the problem and give preliminary results on perfectly representable conditional distributions. In Section~\ref{sec:main:results}, we give our necessary and sufficient condition for perfect representability. In Section~\ref{algorithmic results chapter}, we formulate the problem as a linear program. In Section~\ref{computational complexity chapter}, we study the complexity of solving this linear program and prove that deciding perfect representability is NP-complete. Lastly, in Section~\ref{relations to infor measures section}, we discuss the connections with several information measures and conclude in Section~\ref{concluding remarks section}.

\section{Preliminaries}\label{sec:system:model}

Let $\mathcal{X}$ and $\mathcal{Y}$ denote the supports of $X$ and $Y$, respectively. We write $x\in\mathcal{X}$ and $y\in\mathcal{Y}$ for their respective realizations. We assume, without loss of generality, that $\mathcal{X} = \{1,2,...,k\}$ and $\mathcal{Y} = \{1,2,...,m\}$ for some $k,m \in \mathbb{Z}^+$.

\subsection{Basic Definitions}

Given a functional representation $Z$ of $(X,Y)$ such that $Y=f(X,Z)$, define $W_x:=f(x,Z)$ for each $x\in\mathcal{X}$. The vector
\begin{align}\label{eq: canonical definition equation}
    W = \begin{bmatrix}
f(1,Z) & f(2,Z) & \cdots & f(k,Z)
\end{bmatrix}^T
\end{align}
has the following properties:
\begin{enumerate}
    \item $W$ is independent of $X$.
    \item  With $g(x,w) := w_x$ (i.e. $x$-th entry of the vector $w$), we have $Y = g(X,W)$, consequently, $W$ is also a functional representation of $(X,Y)$.
    \item $W_x\sim p_{Y\mid X=x}$ for each $x\in\mathcal{X}$.
    \item $I(Y;W) = I(Y;Z)$.
\end{enumerate}
The first three properties are trivial, and the last one is shown in Lemma \ref{optimality of canonical FR} below. This observation motivates the following definition which also follows from \cite[Lemma 3.22]{kallenberg2002}.

\begin{definition}For a given $(X,Y) \sim p_{XY}$, a canonical functional representation of $(X,Y)$ is a random vector $[W_1, W_2, ... , W_k]^T$ independent of $X$ such that $W_x \sim p_{Y\mid X=x}$ for every $x\in\mathcal{X}$ and $Y=W_X$ almost surely.
\end{definition}

\begin{remark}\label{remark 1 canonical fr}
    If $W$ is a canonical functional representation for $(X,Y)$, it is also a canonical functional representation for all $(X',Y')$ with conditional distribution $p_{Y|X}$ regardless of the distribution of $X'$.
\end{remark}

\begin{lemma}\label{optimality of canonical FR}
    Given an $(X, Y, Z)$ triple that satisfies (\ref{functional representation restrictions}), there is a canonical functional representation $W$ for the pair $(X,Y)$ that satisfies $I(Y;W) = I(Y;Z)$.
\end{lemma}
See Appendix~\ref{appendix A} for the proof.

Note that Lemma \ref{optimality of canonical FR} implies that if there exists a functional representation $Z$ of $(X,Y) \sim p_{XY}$, then there exists a canonical functional representation $W$ that is as informative as $Z$ about $Y$. Thus, it is sufficient to focus on canonical representations.

\begin{definition}\label{def:perfect representability}
    A conditional distribution $p_{Y|X}$ is said to be perfectly representable if there exists $p_{W|Y}$ such that $W$ is a canonical functional representation for $(X,Y)$. %
\end{definition}

If $p_{XYW} = p_{W|Y} p _{Y|X} p_X$, then $X-Y-W$ is a Markov chain. If $W$ is also a canonical functional representation, then $I(Y;W)$ satisfies (\ref{trivial upper bound}) with equality. From Remark \ref{remark 1 canonical fr}, we see that $(X,Y)$ being perfectly representable or not depends on $p_{Y|X}$ only. Hence, our main objective is reduced to investigating the family of conditional distributions $p_{Y|X}$ that are perfectly representable.

\subsection{Singular Channels}
We recall the concept of a singular channel (e.g., \cite{sriramu2024optimalredundancyexactchannel}):

\begin{definition}\label{def: perfect channels}
    A discrete channel $p_{Y|X}$ is called singular if, for all $y \in \mathcal{Y}$ and all $x_1, x_2 \in \mathcal{X}$ with $p_{Y|X}(y|x_1) p_{Y|X}(y|x_2)>0$, we have $p_{Y|X}(y|x_1) = p_{Y|X}(y|x_2)$. Moreover, we call a conditional probability matrix singular if it corresponds to a singular channel.
 
\end{definition}

Just like in \cite{sriramu2024optimalredundancyexactchannel}, the concept of singular channels is important for our problem. As it turns out, singularity is a necessary condition for perfect representability.

\begin{lemma} \label{singularity lemma}
     If a conditional distribution $p_{Y|X}$ is perfectly representable, it is singular. Moreover, if $p_{Y|X}$ is singular and $|\mathcal{X}| = 2$ or $|\mathcal{Y}| = 2$, then $p_{Y|X}$ is perfectly representable.     
\end{lemma}

 According to Lemma \ref{singularity lemma}, singularity is a necessary condition for a channel to be perfectly representable. It is also a sufficient condition when $|\mathcal{X}| = 2$ or $|\mathcal{Y}| = 2$. See Appendix \ref{appendix A} for the proof. However, it is not a sufficient condition in general. 
\begin{example}\label{singular bad matrix}
    For $|\mathcal{X}| = |\mathcal{Y}| = 3$, the channel 
\begin{align} \label{singular bad matrix example}
p_{Y|X} &= \begin{bmatrix}
      1/2 & 1/2 & 0 \\
       1/2 & 0 & 1/2 \\
      0 & 1/2 & 1/2
    \end{bmatrix}
\end{align}
is singular, but not perfectly representable.
\end{example}

It is easy to verify that the matrix in \eqref{singular bad matrix example} in Example~\ref{singular bad matrix} is singular. However, it is not perfectly representable. This can be shown in two ways. First, one may give a direct argument using Definition~\ref{def:perfect representability} and Definition~\ref{def: perfect channels}; we include a formal proof as Lemma \ref{example 1 lemma} in Appendix \ref{appendix A}. Second, one may appeal to Theorem~\ref{main theorem} in Section \ref{sec:main:results} and easily verify that the matrix fails the necessary and sufficient condition for perfect representability stated there.

For a nonsingular channel with binary input, Lemma~\ref{singularity lemma} rules out equality in~\eqref{trivial upper bound} but does not tell us how large $I(Y;Z)$ can be.
When $|\mathcal{X}| = 2$, we can find the functional representation $Z$, for any $p_{XY}$, that achieves the maximum $I(Y;Z)$ and calculate the value of $I(Y;Z)$ in terms of the total variation distance between the conditional distributions.%

\begin{lemma}\label{total variation lemma}
    Given $(X,Y) \sim p_{XY}$ with $|\mathcal{X}| =2$, define
    \begin{align}
r = \text{TV}(p_{Y|X=1} , p_{Y|X=2}).
    \end{align}
    The maximum value of $I(Y;Z)$ among all functional representations $Z$ equals to $H(Y) - r H(X)$. Moreover, $r H(X)$ equals to $I(X;Y)$ if and only if $p_{Y|X}$ is singular.
\end{lemma}

Here, total variation is defined as in~\cite[Sec.~4.1]{levinperes2017}. This is closely connected to the fact that total variation distance gives the probability of error in a maximal coupling. See Appendix \ref{appendix A} for the proof.

\section{Main Results}\label{sec:main:results}

In this section, we give the necessary and sufficient condition for a channel to be perfectly representable. Our approach is to define four basic operations on a singular conditional distribution matrix. Three of these operations preserve the property of being perfectly representable and three of them preserve the property of not being perfectly representable. We show that the set of all perfectly representable channels is exactly the set of channels which could be obtained via perfect-representability preserving operations from the trivial channel that maps every input to the same output.

\subsection{Four Basic Operations} \label{preserving operations}
We can think of the channel $p_{Y|X}$ as a conditional probability matrix $M$ of size $k \times m$ where $k = |\mathcal{X}|$ and $m = |\mathcal{Y}|$. Its entries are $M[x,y]=p_{Y\mid X}(y\mid x)$, so the column indexed by $y$ is $M[:,y]=p_{Y\mid X}(y\mid\cdot)$.
By Lemma \ref{singularity lemma}, all perfectly representable matrices must be singular. Since singularity is easy to check, we focus on singular matrices from now on.
\begin{definition} 
     For a singular conditional probability matrix $M$, the unique non-zero value of the non-zero entries in a column is called the ``value'' of that column. Moreover, a column divided by its value $\beta$ is called the ``pattern'' of that column. So, the pattern of column $y$ is the binary vector $M[:,y] / \beta$.
\end{definition}

In Example \ref{singular bad matrix}, the first column of the singular matrix has value $\beta = 1/2$ and pattern $[1,1,0]^T$.

 We define four operations on singular matrices. Note that it is straightforward to see that the result of these operations will be singular matrices as well. Let $\mathcal{S}$ denote the set of all singular matrices and $\mathcal{S}_{k,m}$ denote the set of singular matrices of size $k \times m$.

\subsubsection{Pattern preserving split (PPS)}
Let $\alpha \in (0,1)$ and $l \in \{1,2,...,m\}$. The pattern preserving split operation $A_{l,\alpha}: \mathcal{S}_{k,m} \rightarrow \mathcal{S}_{k,m+1}$ is the operation that takes a singular matrix $M$ and splits the $l$-th column into two columns (that will correspond to $l$-th and $(l+1)$-th columns in the resulting matrix $A_{l,\alpha}(M)$) that are $\alpha$ and $(1 - \alpha)$ times the original column. %

\begin{example} \label{PPS example}
    Let \begin{align}
    M =  \begin{bmatrix}
      1-\beta & \beta & 0 & 0 & 0 \\
       0 & \beta & 1-\beta & 0 & 0 \\
    1-\beta & 0 & 0 & \beta / 2 & \beta / 2
    \end{bmatrix}, 
\end{align}
then, \begin{align}
    A_{2,1/2}(M) = \begin{bmatrix}
      1-\beta & \beta/2 & \beta/2 & 0 & 0 & 0 \\
       0 & \beta/2 & \beta/2 & 1-\beta & 0 & 0 \\
    1-\beta & 0 & 0 & 0 & \beta / 2 & \beta / 2
    \end{bmatrix}.
\end{align}
The pattern preserving split operation is applied to the second column of $M$ with the splitting ratio of 1/2.
\end{example}

\subsubsection{Pattern preserving merge (PPM)}
Let $l_1$ and $l_2$ be distinct indices in $[1:m+1]$.
The pattern preserving merge operation $B_{l_1,l_2}$ is the operation that takes a singular matrix $M$ and merges $l_1$-th and $l_2$-th columns (assuming that they have the same pattern). The new column will correspond to $l := \min\{l_1,l_2\}$-th column in the resulting matrix $B_{l_1,l_2}(M)$ that equals to the sum of these two columns (i.e. $B_{l_1,l_2}(M)[x,l] = M[x,l_1] + M[x,l_2]$ for all $x\in\mathcal{X}$). Then, it shifts the columns on the right side of $\max\{l_1,l_2\}$-th column to the left and leaves the remaining columns unchanged to get the resulting matrix.
Thus, $B_{l_1,l_2}$ takes as input a singular $k\times(m+1)$ matrix whose $l_1$-th and $l_2$-th columns have the same pattern and returns a singular $k\times m$ matrix.

\begin{example}
    For the same $M$ as in Example \ref{PPS example},  \begin{align}
        B_{4,5}(M) = \begin{bmatrix}
      1-\beta & \beta & 0 & 0 \\
       0 & \beta & 1-\beta & 0  \\
    1-\beta & 0 & 0 & \beta 
    \end{bmatrix}.
    \end{align}The pattern preserving merge operation is applied to the fourth and fifth columns of $M$.
\end{example}

\subsubsection{Value preserving split (VPS)}

Let $l \in [1:m]$ and $\vec{b} \in \{0,1 \}^k$. The value preserving split operation $C_{l,\vec{b}}: \mathcal{S}_{k,m} \rightarrow \mathcal{S}_{k,m+1}$ is the operation that takes a singular matrix and splits the $l$-th column (whose value is $\beta \in (0,1]$ and whose pattern is $\mathcal{\vec{p}} \in \{ 0,1 \}^{k}$) into 2 columns (that will correspond to $l$-th and $(l+1)$-th columns in the resulting matrix $C_{l,\vec{b}}(M)$). Both columns of the new matrix will have value $\beta$ and patterns $\mathcal{\vec{p}}_1 \in \{ 0,1 \}^{k}$ and $\mathcal{\vec{p}}_2 \in \{ 0,1 \}^{k}$ respectively where $\mathcal{\vec{p}}_1 = \mathcal{\vec{p}} \cdot \vec{b}$ and $\mathcal{\vec{p}}_2 = \mathcal{\vec{p}} \cdot (\mathbf{1}_k - \vec{b})$. Choose $\vec{b}$ so that both resulting patterns are nonzero. Note that $\cdot$ represents elementwise multiplication and $\mathbf{1}_k$ represents the vector of size $k$ all whose entries are 1.

\begin{example}
    For the same $M$ as in Example \ref{PPS example}, \begin{align}
        C_{1,[1,1,0]^T}(M) = \begin{bmatrix}
      1-\beta & 0 & \beta & 0 & 0 & 0 \\
       0 & 0 &  \beta& 1-\beta & 0 & 0 \\
    0 & 1-\beta & 0 & 0 & \beta / 2 & \beta / 2
    \end{bmatrix}.
    \end{align} The value preserving split operation is applied to the first column of $M$ with the splitting vector of $[1,1,0]^T$.
\end{example}

\subsubsection{Value preserving merge (VPM)}

Let $l_1$ and $l_2$ be distinct indices in $[1:m+1]$. Suppose that the $l_1$-th and $l_2$-th columns of $M$ have the same value $\beta \in (0,1]$ and patterns $\mathcal{\vec{p}}_1 \in \{ 0,1 \}^{k}$ and $\mathcal{\vec{p}}_2 \in \{ 0,1 \}^{k}$ that satisfy $\mathcal{\vec{p}}_1 + \mathcal{\vec{p}}_2 \in \{ 0,1 \}^{k}$.

The value preserving merge (VPM) operation $D_{l_1,l_2}$ is the operation that takes a singular matrix $M$ and merges $l_1$-th and $l_2$-th columns into a single column with the pattern $\mathcal{\vec{p}}_1 + \mathcal{\vec{p}}_2$ and the value $\beta$ (that will correspond to $\min\{l_1,l_2\}$-th column in the resulting matrix $D_{l_1,l_2}(M)$).
Thus, $D_{l_1,l_2}$ takes as input a singular $k\times(m+1)$ matrix whose $l_1$-th and $l_2$-th columns have the same value and disjoint supports and returns a singular $k\times m$ matrix.

\begin{example}
    For the same $M$ as in Example \ref{PPS example}, \begin{align}
    D_{1,3}(M) = \begin{bmatrix}
      1-\beta & \beta  & 0 & 0 \\
       1-\beta & \beta  & 0 & 0 \\
    1-\beta & 0 & \beta / 2 & \beta / 2
    \end{bmatrix} .
\end{align} The value preserving merge operation is applied to the first and third columns of $M$.
\end{example}

We explain how the order of these operations can be manipulated in the following lemma.

\begin{lemma}\label{reordering operations lemma}
    For any sequence of PPS, VPS, PPM operations, there exists an equivalent cascade that consists of a sequence of PPS operations, followed by a sequence of VPS operations, and finally a sequence of PPM operations.
\end{lemma}
See Appendix~\ref{appendix B} for the proof of Lemma~\ref{reordering operations lemma}. We will use this lemma later in the proof of Theorem~\ref{LP algorithm theorem}.

\begin{lemma} \label{preserving operations lemma}
    The result of applying any one of PPM, PPS, and VPS operations described in \cref{preserving operations} to a perfectly representable channel will also be a perfectly representable channel. 
\end{lemma}

 See Appendix~\ref{appendix B} for the proofs of Lemmas~\ref{preserving operations lemma} and~\ref{preserving operations lemma not}. Since PPM and PPS operations are inverses of each other, the result of applying PPM and PPS operations to a channel that is not perfectly representable will also be a channel that is not perfectly representable. Here, by `inverse' we mean that, with a suitable choice of parameters, each operation can undo the effect of its counterpart on a matrix. Moreover, since VPM and VPS operations are inverses of each other, the result of applying VPM operation to a channel that is not perfectly representable will also be a channel that is not perfectly representable. Thus, Lemma \ref{preserving operations lemma not} is a direct result of Lemma \ref{preserving operations lemma}.

\begin{lemma} \label{preserving operations lemma not}
    The result of applying any one of PPM, PPS, and VPM operations described in \cref{preserving operations} to a channel that is not perfectly representable will also be a channel that is not perfectly representable.     
\end{lemma}

\subsection{Perfect Representability} \label{perfect representability}

We are now ready to find the family of 
 channels that are perfectly representable. We start with a channel $p_{Y|X}$ that maps every $x \in \mathcal{X}$ to the same $a \in \mathcal{Y}$. This channel has a conditional probability matrix which is the all ones vector of length $k$ and is clearly perfectly representable (choose $Z=Y$). By Lemma \ref{preserving operations lemma}, all matrices that we obtain from this matrix with the repeated use of pattern preserving split, pattern preserving merge, and value preserving split operations are perfectly representable. Finally, we will prove that these conditional probability matrices are the only perfectly representable conditional probability matrices.

\begin{theorem} \label{main theorem}
    A singular $k \times m$ conditional probability matrix $M$ is perfectly representable if and only if it is possible to obtain the vector $\mathbf{1}_k$ from $M$ via a repeated use of PPM, PPS, and VPM operations defined in \cref{preserving operations}. 
\end{theorem} 

Note that the ``if'' direction follows directly from Lemmas~\ref{preserving operations lemma} and~\ref{preserving operations lemma not}. See Appendix~\ref{appendix B} for the proof.

\begin{example}\label{reducing example}
The following example illustrates how a representable matrix can be reduced to $\mathbf{1}_k$ using the operations in Theorem~\ref{main theorem}. Consider
\begin{align}
M
=
\begin{bmatrix}
1/3 & 1/3 & 0 & 0 & 1/3\\
1/3 & 0 & 1/3 & 0 & 1/3\\
0 & 1/3 & 0 & 1/3 & 1/3
\end{bmatrix}.
\end{align}
Equivalently, writing the columns in terms of their values and patterns,
\begin{align}
M
=
\left[
\tfrac13\!\begin{pmatrix}1\\1\\0\end{pmatrix},
\tfrac13\!\begin{pmatrix}1\\0\\1\end{pmatrix},
\tfrac13\!\begin{pmatrix}0\\1\\0\end{pmatrix},
\tfrac13\!\begin{pmatrix}0\\0\\1\end{pmatrix},
\tfrac13\!\begin{pmatrix}1\\1\\1\end{pmatrix}
\right].
\end{align}
We now reduce $M$ to $\mathbf{1}_3$.

First, apply VPM to the second and third columns:
\begin{align}
\left[
\tfrac13\!\begin{pmatrix}1\\1\\0\end{pmatrix},
\tfrac13\!\begin{pmatrix}1\\0\\1\end{pmatrix},
\tfrac13\!\begin{pmatrix}0\\1\\0\end{pmatrix},
\tfrac13\!\begin{pmatrix}0\\0\\1\end{pmatrix},
\tfrac13\!\begin{pmatrix}1\\1\\1\end{pmatrix}
\right]\\
\xrightarrow[\text{vpm}]{} 
\left[
\tfrac13\!\begin{pmatrix}1\\1\\0\end{pmatrix},
\tfrac13\!\begin{pmatrix}1\\1\\1\end{pmatrix},
\tfrac13\!\begin{pmatrix}0\\0\\1\end{pmatrix},
\tfrac13\!\begin{pmatrix}1\\1\\1\end{pmatrix}
\right].
\end{align}

Next, apply PPM to the two columns with pattern $\mathbf{1}_3$:
\begin{align}
\left[
\tfrac13\!\begin{pmatrix}1\\1\\0\end{pmatrix},
\tfrac13\!\begin{pmatrix}1\\1\\1\end{pmatrix},
\tfrac13\!\begin{pmatrix}0\\0\\1\end{pmatrix},
\tfrac13\!\begin{pmatrix}1\\1\\1\end{pmatrix}
\right]\\
\xrightarrow[\text{ppm}]{} 
\left[
\tfrac13\!\begin{pmatrix}1\\1\\0\end{pmatrix},
\tfrac13\!\begin{pmatrix}0\\0\\1\end{pmatrix},
\tfrac23\!\begin{pmatrix}1\\1\\1\end{pmatrix}
\right].
\end{align}

Then, apply VPM to the first two columns:
\begin{align}
\left[
\tfrac13\!\begin{pmatrix}1\\1\\0\end{pmatrix},
\tfrac13\!\begin{pmatrix}0\\0\\1\end{pmatrix},
\tfrac23\!\begin{pmatrix}1\\1\\1\end{pmatrix}
\right]
\xrightarrow[\text{vpm}]{} 
\left[
\tfrac13\!\begin{pmatrix}1\\1\\1\end{pmatrix},
\tfrac23\!\begin{pmatrix}1\\1\\1\end{pmatrix}
\right].
\end{align}

Finally, apply PPM:
\begin{align}
\left[
\tfrac13\!\begin{pmatrix}1\\1\\1\end{pmatrix},
\tfrac23\!\begin{pmatrix}1\\1\\1\end{pmatrix}
\right]
\xrightarrow[\text{ppm}]{} 
\begin{pmatrix}1\\1\\1\end{pmatrix}
=
\mathbf{1}_3.
\end{align}

Hence, by Theorem~\ref{main theorem}, the matrix $M$ is perfectly representable.
\end{example}

\begin{remark}
By the equivalences noted in the introduction, Theorem~\ref{main theorem} also characterizes exactly when
\begin{itemize}
    \item $H_K(G_{YX},Y)=H(Y|X)$; see~\cite{Korner1973,Alon^2,LiElGamal2018EGW};
    \item $G_{\mathrm{PF}}(H(Y|X),X\!\to\!Y)=0$; see~\cite{privacy_funnel}.
\end{itemize}
Equivalently, it characterizes exactly when the redundancy term in channel synthesis is zero; see~\cite{sriramu2024optimalredundancyexactchannel}. The graph entropy and privacy funnel equivalences are explained in more detail in Section~\ref{relations to infor measures section}.
\end{remark}

\section{Algorithmic Results}\label{algorithmic results chapter}
In this section, we will focus on the problem of checking if a conditional distribution $p_{Y|X}$ is perfectly representable.

Note that checking whether a matrix is a singular conditional probability matrix or not can be done in linear time. Thus, we can assume that we are working with singular conditional probability matrices. A matrix $M$ is obtainable from $\mathbf{1}_k$ by PPS, PPM, VPS operations if and only if the matrix $M'$ that is obtained from $M$ by merging all columns with the same pattern is obtainable from $\mathbf{1}_k$ by PPS, PPM, VPS operations. Therefore, we can assume that each column of $M$ has a distinct pattern. That is, it has $m \leq 2^k-1$ columns with distinct patterns $\vec{p}_1, \vec{p}_2, ... , \vec{p}_m$ with values $v_1, v_2, ...,v_m$ in $[0,1]$. %
Thus, our singular conditional probability matrix could be solely defined by the vector of values $\vec{v}$ and patterns $\vec{p}_1, \vec{p}_2, ... , \vec{p}_m$.

Consider all possible sets of nonzero binary vectors of length $k$ that add up to $\mathbf{1}_k$. Note that the addition here is in real-number arithmetic, not modulo-$2$ arithmetic. Denote these sets by
\begin{align}
    \widehat{S}_1,\widehat{S}_2,\ldots,\widehat{S}_{B(k)},
\end{align}
where $B(k)$ is the $k$-th Bell number, i.e., the number of set partitions of a $k$-element set~\cite{bell_number}.

\begin{example}
For $k=3$, we have $B(3)=5$, and the sets are
\begin{align}
    \widehat{S}_1
    &=
    \{[1,1,1]^T\},\\
    \widehat{S}_2
    &=
    \{[1,1,0]^T,[0,0,1]^T\},\\
    &\hspace{1em}\ldots,\\
    \widehat{S}_5
    &=
    \{[1,0,0]^T,[0,1,0]^T,[0,0,1]^T\}.
\end{align}
\end{example}

Let
\begin{align}
    \widehat{\vec{p}}_1,\widehat{\vec{p}}_2,\ldots,
    \widehat{\vec{p}}_{2^k-1}
\end{align}
be all nonzero binary vectors of length $k$. Without loss of generality, order them such that
\begin{align}
    \widehat{\vec{p}}_i=\vec{p}_i,
    \qquad i\in\{1,2,\ldots,m\},
\end{align}
where $\vec{p}_1,\ldots,\vec{p}_m$ are the patterns occurring in $M$. In the linear program (LP) we will formulate below, $i$ indexes patterns and $j$ indexes partitions.

Define the extended value vector
\begin{align}
    \widehat{\vec{v}}
    =
    (\widehat{v}_1,\ldots,\widehat{v}_{2^k-1})^T
\end{align}
by
\begin{align}
    \widehat{v}_i
    =
    \begin{cases}
        v_i,
        & i\in\{1,2,\ldots,m\},\\
        0,
        & i\in\{m+1,\ldots,2^k-1\}.
    \end{cases}
\end{align}
Thus, the entries of $\widehat{\vec{v}}$ corresponding to patterns that do not occur in $M$ are zero.

Now, define
\begin{align}
    \widehat{H}
    \in
    \{0,1\}^{(2^k-1)\times B(k)}
\end{align}
by
\begin{align}
    \widehat{H}_{ij}
    =
    \mathds{1}
    \{
    \widehat{\vec{p}}_i\in\widehat{S}_j
    \}.
\end{align}
We now restrict our attention to the partitions that use only patterns occurring in $M$. Define
\begin{align}
    \mathcal{J}
    :=
    \left\{
    j\in\{1,2,\ldots,B(k)\}:
    \widehat{S}_j
    \subseteq
    \{\vec{p}_1,\vec{p}_2,\ldots,\vec{p}_m\}
    \right\}.
\end{align}
Thus, $\mathcal{J}$ indexes the partitions that use only patterns occurring in $M$. Let
\begin{align}
    R:=|\mathcal{J}|\leq B(k),
\end{align}
and, after reindexing, denote the corresponding sets by
\begin{align}
    S_1,S_2,\ldots,S_R.
\end{align}

Define
\begin{align}
    H\in\{0,1\}^{m\times R}
\end{align}
by
\begin{align}
    H_{ij}
    =
    \mathds{1}\{\vec{p}_i\in S_j\}.
\end{align}
Equivalently, $H$ is obtained from $\widehat{H}$ by retaining the first $m$ rows and the columns indexed by $\mathcal{J}$.

\begin{example}
For $k=3$, suppose that the only patterns occurring in $M$ are
\begin{align*}
    \vec{p}_1&=[1,0,0]^T, \qquad \vec{p}_2=[0,1,0]^T,\\
    \vec{p}_3&=[0,0,1]^T, \qquad \vec{p}_4=[1,1,0]^T.
\end{align*}
Order the remaining patterns as $[1,0,1]^T$, $[0,1,1]^T$, and $[1,1,1]^T$.
Of the $B(3)=5$ partitions, $\widehat{S}_2=\{\vec{p}_3,\vec{p}_4\}$ and $\widehat{S}_5=\{\vec{p}_1,\vec{p}_2,\vec{p}_3\}$ are the only ones formed from these four patterns.
Hence, $R=2$, and $H$ is obtained by retaining the first four rows of $\widehat{H}$ and its second and fifth columns.
\[
    \widehat{H}=
    \begin{pmatrix}
        0&0&0&1&1\\
        0&0&1&0&1\\
        0&1&0&0&1\\
        0&1&0&0&0\\
        0&0&1&0&0\\
        0&0&0&1&0\\
        1&0&0&0&0
    \end{pmatrix},
    \qquad
    H=
    \begin{pmatrix}
        0&1\\
        0&1\\
        1&1\\
        1&0
    \end{pmatrix}.
\]
\end{example}

\begin{theorem}\label{LP algorithm theorem}
Let $\vec{v}$ and $H$ be the vector and matrix obtained from $M$ by the above procedure. Then $M$ is perfectly representable if and only if $H\vec{x}=\vec{v}$ has a nonnegative solution. Equivalently, $M$ is perfectly representable if and only if
\begin{align}
    \max_{\substack{\vec{x}\geq0\\H\vec{x}\leq\vec{v}}}
    \vec{1}_R\cdot\vec{x}
    =
    1.
\end{align}
\end{theorem}
See Appendix~\ref{appendix C} for the proof. After checking that $M$ is a singular conditional probability matrix, we can determine whether $M$ is perfectly representable by solving this linear program.

%% file: computational_complexity_v2.tex
\subsection{Solving the Dual LP via a Separation Oracle}
\label{dual separation subsection}

The dual of the LP in Theorem~\ref{LP algorithm theorem} has $m$ variables and gives the following equivalent condition.

\begin{corollary}\label{Dual LP corollary}
A matrix $M$ is perfectly representable if and only if the corresponding vector $\vec{v}$ and the matrix $H$ obtained by the procedure in Section~\ref{algorithmic results chapter} satisfy
\begin{align}
    \min_{\substack{\vec{y}\geq0\\H^T\vec{y}\geq\vec{1}_R}}
    \vec{v}\cdot\vec{y}
    =
    1.
    \label{reduced dual LP}
\end{align}
\end{corollary}

Here, the variables are
\begin{align}
    \vec{y}=(y_1,y_2,\ldots,y_m),
\end{align}
so the dual has
\begin{align}
    m\leq2^k-1
\end{align}
variables and
\begin{align}
    R\leq B(k)
\end{align}
constraints.

Since the Bell numbers grow super-exponentially
in $k$~\cite{bellnumber_Onotation}, explicitly constructing
all partition constraints can be prohibitively expensive.

\addvspace{6pt}
\noindent\begin{minipage}{\columnwidth}
\begin{theorem}\label{dual complexity theorem}
Let $M$ be a rational singular conditional probability matrix with $k$ rows and $m$ columns with distinct occurring patterns, and let $L$ be its binary encoding length. The optimal value of the dual LP in Corollary~\ref{Dual LP corollary} can be computed exactly in $2^k\operatorname{poly}(m,L)$ time using a separation oracle, without listing all partition constraints.
\end{theorem}
\end{minipage}\par
\vspace{6pt}
For a candidate vector $\vec{y}\geq0$, give each occurring pattern $\vec{p}_i$ weight $y_i$.
The separation oracle computes the minimum weight of a partition of $[k]$ into supports of these patterns.
If this minimum is less than $1$, a minimizing partition gives a violated constraint.
Otherwise, all partition constraints are satisfied.
We compute the minimum by dynamic programming over the $2^k$ subsets of $[k]$, scanning at most $m$ patterns at each state.
The ellipsoid method then gives exact optimization using $\operatorname{poly}(m,L)$ separation calls and polynomial additional computation.
See Appendix~\ref{dual complexity proof subsection} for the proof.

Thus, the dual optimum can be computed exactly and compared
with one in polynomial time for fixed $k$, without explicitly
listing all partition constraints. When $k$ is part of the
input, exact optimization remains NP-hard, as shown in
Appendix~\ref{RXC3 appendix subsection}. This does not establish that the
dependence $2^k$ is optimal.

\subsection{NP-Completeness of Deciding Perfect Representability}
\label{perfect representability hardness subsection}

The preceding optimization hardness does not by itself settle the complexity of deciding perfect representability, because this decision problem only asks whether the optimal value equals one. We therefore give a separate reduction and prove membership in NP.

\begin{theorem}
\label{perfect representability NP-hard theorem}
Given a rational singular conditional probability matrix encoded in binary, deciding whether it is perfectly representable is NP-complete.
\end{theorem}

In the proof in Appendix~\ref{decision complexity proof subsection}, we construct a singular conditional probability matrix $M(G)$ from a finite simple graph $G$ in polynomial time.
A proper 3-coloring of $G$ and its two cyclic relabelings give three exact partitions of the rows of $M(G)$, each of weight $1/3$. Conversely, any perfect representation of $M(G)$ yields an exact partition and hence a proper 3-coloring of $G$.
Since graph 3-colorability is NP-complete~\cite{garey_johnson_1979}, the reduction gives NP-hardness even for matrices with three nonzero entries of value $1/3$ per row.
For every perfectly representable rational singular conditional probability matrix, a list of at most $m$ exact partitions and their rational weights gives a polynomial-size certificate that can be checked in polynomial time. This establishes membership in NP; see Appendix~\ref{decision complexity proof subsection} for the details.
Thus, deciding perfect representability is NP-complete for rational singular conditional probability matrices encoded in binary, even when every row has exactly three nonzero entries, all equal to $1/3$.

%% file: common_info.tex
In Theorem~\ref{main theorem}, we found a necessary and sufficient condition for perfect representability. This characterization naturally gives rise to the following corollaries on K{\"o}rner graph entropy, the privacy funnel, excess functional information, and Wyner's common information. The proofs are given in Appendix~\ref{appendix E}.

\subsection{Körner Graph Entropy}
Let $G=(V,E)$ be a graph and let $X\sim P$ be a random vertex in $V$.
Let $\mathcal{I}(G)$ denote the family of independent sets of $G$.
The K{\"o}rner graph entropy of $G$ under $P$~\cite{Korner1973,Alon^2} is
\[
H(G,P)\;:=\;\min I(X;Y),
\]
where the minimum is over all random variables $Y\in\mathcal{I}(G)$ such that
$X\in Y$ almost surely.

Now, let $G_{YX}$ be the graph with vertex set $\mathcal{Y}$ and an edge between distinct vertices $y_1$ and $y_2$ if $p(x,y_1) p(x,y_2) >0$ for some $x$. We denote its K{\"o}rner graph entropy under $p_Y$ by $H_K(G_{YX},Y)$.

\begin{corollary}\label{cor:graph-entropy}
The channel $p_{Y|X}$ is perfectly representable if and only if $H_K(G_{YX},Y) = H(Y|X)$.
\end{corollary}

\subsection{Privacy Funnel}
Let $(X,Y)\sim P_{XY}$ be a pair of random variables, where $Y$ denotes a public variable that can be disclosed, and $X$ denotes a private variable that should be protected.
A privacy mechanism is defined by a randomized mapping $P_{U|Y}$ producing a released representation $U$, such that the Markov chain $X - Y - U$ holds.

The privacy funnel~\cite{privacy_funnel} quantifies the optimal trade-off between preserving information about the public variable $Y$ and limiting the information leakage about the private variable $X$. For a given disclosure level $0\le t\le H(Y)$, it is defined as
\begin{align*}
    G_{\mathrm{PF}}(t, X \!\to\! Y)
\;=\;
\min_{U:\, X-Y-U,\; I(Y;U)\ge t} I(X;U).
\end{align*}

That is, among all released representations $U$ that retain at least $t$ bits of information about the public variable $Y$, the privacy funnel characterizes the minimum possible information leakage about the private variable $X$.

\begin{corollary}\label{cor:privacy-funnel}
The channel $p_{Y|X}$ is perfectly representable if and only if $G_{\mathrm{PF}}\!\big(H(Y|X),\, X \!\to\! Y\big) = 0$.
\end{corollary}

\subsection{Excess Functional Information}
Given random variables $(X,Y) \sim p_{XY}$, the excess functional information~\cite{SFRL} is defined as
\begin{align*}
    \Psi(X \rightarrow Y) = \min_{U: I(U;X)=0} H(Y|U) - I(X;Y).
\end{align*}

\begin{corollary}\label{cor:excess-functional}
The channel $p_{Y|X}$ is perfectly representable if and only if $\Psi(X \rightarrow Y) = 0$.
\end{corollary}

\subsection{Wyner's Common Information}
We denote Wyner's common information~\cite{wyner1975common} by $C(X,Y)$, defined as
\[
C(X,Y) = \inf_{p_{W|XY}:X-W-Y} I(XY;W).
\]
Note that $C(X,Y) \geq I(X;Y)$~\cite{wyner1975common}.

\begin{corollary}\label{cor:wyner}
If $C(X,Y) = I(X;Y)$, then $p_{Y|X}$ is perfectly representable. However, perfect representability can also hold when $C(X,Y) > I(X;Y)$.
\end{corollary}

%% file: concluding_remarks.tex
We characterized perfectly representable conditional distributions through elementary operations on singular conditional probability matrices and gave an equivalent linear-programming formulation. Using a subset-dynamic-programming separation oracle, the compact dual LP can be solved in $2^k\operatorname{poly}(m,L)$ time and hence in polynomial time for fixed $k=|\mathcal X|$. In contrast, when $k$ is part of the input, exact optimization of the compact LP is NP-hard, and deciding perfect representability of a rational input matrix is NP-complete. These results provide both a structural characterization of perfect functional representations and a computational distinction between fixed-alphabet and general instances.

%% file: appendixA.tex
\begin{proof}[Proof of Lemma~\ref{optimality of canonical FR}]
    Choose the random variable $W$ as in equation \eqref{eq: canonical definition equation}. Then, $W$ is a function of $Z$. By applying the equality in (\ref{trivial upper bound}) to both $Z$ and $W$, we get
\begin{align}
I(Y;Z) - I(Y;W) = I(X;W|Y) - I(X;Z|Y).\label{eq. in proof of lemma 1}
\end{align}
 Since $W$ is a function of $Z$, the left hand side of (\ref{eq. in proof of lemma 1}) is non-negative and the right hand side of (\ref{eq. in proof of lemma 1}) is non-positive. Consequently, both sides are equal to zero.
\end{proof}

\begin{proof}[Proof of Lemma \ref{singularity lemma}]
We have three statements to prove: 
\\
\textbf{1) Perfectly representable channels are singular.}

    Assume that $p_{Y|X}$ is perfectly representable. Then there exists a random vector $W=[W_1,\ldots,W_k]^T$ such that $I(X;W)=0$, $Y=W_X$, and $X-Y-W$.

Fix $x \in \mathcal X$, $y \in \mathcal Y$, and $w=(w_1,\ldots,w_k) \in \mathcal Y^k$. Since $I(X;W) = 0$ and $Y=W_X$, we have
\begin{align}
\Pr[X=x,W=w,Y=y]
&= p_X(x)\,p_W(w)\,\mathds{1}\{w_x=y\}. \label{eq:singularity_lemma_eq_1}
\end{align}
On the other hand, since $X-Y-W$,
\begin{align}
\Pr[X=x,W=w,Y=y]
&= p_X(x)\,p_{Y|X}(y|x)\,p_{W|Y}(w|y). \label{eq:singularity_lemma_eq_2}
\end{align}
Equating \eqref{eq:singularity_lemma_eq_1} and \eqref{eq:singularity_lemma_eq_2}, and using $p_X(x)>0$ for all $x \in \mathcal X$, we obtain
\begin{align}
p_W(w)\,\mathds{1}\{w_x=y\}
= p_{Y|X}(y|x)\,p_{W|Y}(w|y). \label{eq:singularity_lemma_eq_3}
\end{align}

Now fix $y \in \mathcal Y$ and suppose that $p_{Y|X}(y|x_1)>0$ and $p_{Y|X}(y|x_2)>0$ for some $x_1,x_2 \in \mathcal X$.
Choose any $w$ such that $p_{W|Y}(w|y)>0$; such a $w$ exists because $p_{W|Y}(\cdot|y)$ is a probability distribution.
Then \eqref{eq:singularity_lemma_eq_3} implies, for $i=1,2$,
\begin{align}
p_W(w)\,\mathds{1}\{w_{x_i}=y\}
= p_{Y|X}(y|x_i)\,p_{W|Y}(w|y) > 0.    
\end{align}

Hence $\mathds{1}\{w_{x_i}=y\}=1$, so $w_{x_1}=w_{x_2}=y$. Substituting back into \eqref{eq:singularity_lemma_eq_3} gives
\begin{align}
    p_W(w)=p_{Y|X}(y|x_i)\,p_{W|Y}(w|y), \qquad i=1,2.
\end{align}

Therefore
\begin{align}
    p_{Y|X}(y|x_1)=p_{Y|X}(y|x_2).
\end{align}
Since this holds for every $y$ and every pair $x_1,x_2$ with
$p_{Y|X}(y|x_1)>0 \text{ and } p_{Y|X}(y|x_2)>0$, the channel $p_{Y|X}$ is singular.
\\
\textbf{2) Singular channels with $|\mathcal{X}| = 2$ are perfectly representable.}

Without loss of generality, assume $\mathcal X=\{1,2\}$ and let
\begin{align}
    p_x(y):=p_{Y|X}(y|x), \qquad x\in\{1,2\}.
\end{align}

Since the channel is singular, the alphabet $\mathcal Y$ can be partitioned into
\begin{align}
\mathcal A &:= \{y \in \mathcal Y : p_1(y)=p_2(y)>0\},\\
\mathcal B &:= \{y \in \mathcal Y : p_1(y)>0,\; p_2(y)=0\},\\
\mathcal C &:= \{y \in \mathcal Y : p_2(y)>0,\; p_1(y)=0\}.
\end{align}
Let
\begin{align}
    \alpha := \sum_{y \in \mathcal A} p_1(y)
        = \sum_{y \in \mathcal A} p_2(y).
\end{align}

If $\alpha=1$, then $p_1=p_2$, and the choice $W=(U,U)$ with
$U\sim p_1$ is already a perfect functional representation.
If $\alpha=0$, omit the common component and let
\begin{align}
    W=(V_1,V_2)
\end{align}
almost surely, where $V_1\sim p_1$ and $V_2\sim p_2$.
The argument below then applies directly. Hence, for the
remaining case, assume
\begin{align}
    0<\alpha<1.
\end{align}

Define probability distributions
\begin{align}
    q_{\mathcal A}(y):=\frac{p_1(y)}{\alpha}, \quad y\in\mathcal A,
\\
q_{\mathcal B}(y):=\frac{p_1(y)}{1-\alpha}, \quad y\in\mathcal B,
\\
q_{\mathcal C}(y):=\frac{p_2(y)}{1-\alpha}, \quad y\in\mathcal C.
\end{align}

Now construct a random vector $W=(W_1,W_2)$, independent of $X$, as follows:
\begin{align}
W=
\begin{cases}
(U,U), \text{with probability } \alpha \\
(V_1,V_2), \text{with probability } 1-\alpha
\end{cases}
\end{align}
where $V_1$ and $V_2$ are independent and $U\sim q_{\mathcal A}$, $V_1\sim q_{\mathcal B}$, and $V_2\sim q_{\mathcal C}$.

Let $Y:=W_X$. By construction,
\begin{align}
    W_1 \sim p_1, \qquad W_2 \sim p_2,
\end{align}
hence $p_{Y|X}(y|x)=p_x(y)$ for $x\in\{1,2\}$. Thus $W$ is a canonical functional representation of $(X,Y)$.

It remains to verify that $X-Y-W$. Fix any
$y\in\mathcal Y$ with $p_Y(y)>0$. If $y\in\mathcal A$, then
conditioning on $Y=y$ implies
\begin{align}
    W=(y,y)
\end{align}
almost surely. Hence, $W$ is deterministic conditioned on
$Y=y$ and is therefore independent of $X$ conditioned on
$Y=y$.

If $y\in\mathcal B$, then $p_{Y|X}(y|2)=0$, so $X=1$
almost surely conditioned on $Y=y$. Similarly, if
$y\in\mathcal C$, then $X=2$ almost surely conditioned on
$Y=y$. Hence, in either case, $X$ is deterministic conditioned
on $Y=y$ and is therefore independent of $W$ conditioned on
$Y=y$.

Thus, $X$ and $W$ are conditionally independent given every
possible value of $Y$, and consequently
\begin{align}
    X-Y-W.
\end{align}

Hence $W$ is a perfect functional representation, and $p_{Y|X}$ is perfectly representable.
\\
\textbf{3) Singular channels with $|\mathcal{Y}| = 2$ are perfectly representable.}

Without loss of generality, let $\mathcal Y=\{1,2\}$. We have 2 cases:

\begin{itemize}
    \item[\textbf{a)}] Suppose first that there exists $x_0 \in \mathcal X$ such that $0<p_{Y|X}(1|x_0)<1$. Then also $0<p_{Y|X}(2|x_0)<1$.

We claim that in this case all rows of $p_{Y|X}$ are identical. Indeed, fix any $x\in\mathcal X$. If
\begin{align}
    p_{Y|X}(1|x)=0,
\end{align}
then
\begin{align}
    p_{Y|X}(2|x)=1.
\end{align}
Since $p_{Y|X}(2|x_0)>0$, singularity would imply
\begin{align}
    p_{Y|X}(2|x)
    =
    p_{Y|X}(2|x_0),
\end{align}
which is impossible because the left-hand side equals one while the right-hand side is strictly less than one. Similarly, if
\begin{align}
    p_{Y|X}(1|x)=1,
\end{align}
then singularity applied to the output symbol $1$ would imply
\begin{align}
    p_{Y|X}(1|x)
    =
    p_{Y|X}(1|x_0),
\end{align}
which is again impossible. Hence,
\begin{align}
    0<p_{Y|X}(1|x)<1,
    \qquad
    \forall x\in\mathcal X.
\end{align}

By singularity, whenever $p_{Y|X}(1|x)$ and $p_{Y|X}(1|x_0)$ are positive, they must be equal, so
\begin{align}
    p_{Y|X}(1|x)=p_{Y|X}(1|x_0), \qquad \forall x \in \mathcal X.
\end{align}
Therefore,
\begin{align}
    p_{Y|X}(\cdot|x)=p_{Y|X}(\cdot|x_0), \qquad \forall x \in \mathcal X.
\end{align}
Hence, $X$ and $Y$ are independent. Then, $p_{Y|X}$ is perfectly representable as $Y$ is a perfect functional representation for $(X,Y)$.
\item[\textbf{b)}] Assume that there exists no $x_0 \in \mathcal X$ such that $0<p_{Y|X}(1|x_0)<1$. Then, for every $x \in \mathcal X$, $p_{Y|X}(1|x) \in \{0,1\}$, so the channel is deterministic. 

Now, define $f:\mathcal X \to \mathcal Y$ by
\begin{align}
    p_{Y|X}(f(x)|x)=1,\ \  \forall x \in \mathcal X,
\end{align}
and let
\begin{align}
    W=(f(1),\ldots,f(k)).
\end{align}
Then $W$ is deterministic, hence independent of $X$. Also, $ Y=W_X$, and since $W$ is deterministic we trivially have $X-Y-W$. Hence, $W$ is a perfect functional representation.

Therefore, if $p_{Y|X}$ is singular and $|\mathcal Y|=2$, then $p_{Y|X}$ is perfectly representable. 
\end{itemize}

\end{proof}

\begin{lemma}\label{example 1 lemma}
    Singular channels are not necessarily perfectly representable. For instance, for $|\mathcal{X}| = |\mathcal{Y}| = 3$, the channel 
\begin{align} \label{singular bad matrix example appendix}
p_{Y|X} &= \begin{bmatrix}
      1/2 & 1/2 & 0 \\
       1/2 & 0 & 1/2 \\
      0 & 1/2 & 1/2
    \end{bmatrix}
\end{align}
is singular, but not perfectly representable.
\end{lemma}
\begin{proof}
It is easy to verify that the matrix in \eqref{singular bad matrix example appendix} is singular. However, it is not perfectly representable.
To see that the channel $p_{Y|X}$ is not perfectly representable, assume for contradiction that it is perfectly representable, and let $X \sim \mathrm{Unif}\{1,2,3\}$. Then there exists a perfect canonical functional representation
\begin{align}
    W=(W_1,W_2,W_3)
\end{align}
such that
\begin{align}
    W_1 \sim \mathrm{Unif}\{1,2\}, 
    W_2 \sim \mathrm{Unif}\{1,3\}, 
    W_3 \sim \mathrm{Unif}\{2,3\},
\end{align}
and
\begin{align}
    X-Y-W.
\end{align}

Since $\Pr[W_1=1]=1/2$, there exists $w=(w_1,w_2,w_3)$ in the support of $W$ such that $w_1=1$.
For such a $w$, we have
\begin{align}
    \Pr[Y=1,W=w] \geq \Pr[X=1,W=w] > 0,
\end{align}
because $Y=W_X$ and $w_1=1$. Hence
\begin{align}
    \Pr[W=w \mid Y=1] > 0.
\end{align}
Now, since $X$ is uniform and $Y=1$ can arise only from $X=1$ or $X=2$, we have
\begin{align}
    \Pr[X=1 \mid Y=1]=\Pr[X=2 \mid Y=1]=\frac12.
\end{align}
Because $X-Y-W$, conditioning further on $W=w$ cannot change this conditional distribution, so
\begin{align}
    \Pr[X=1 \mid Y=1,W=w]=\frac12.
\end{align}
If $w_2 \neq 1$, then the event $\{Y=1,W=w\}$ forces $X=1$, and therefore
\begin{align}
    \Pr[X=1 \mid Y=1,W=w]=1,
\end{align}
which is a contradiction. Thus we must have
\begin{align}
    w_2=1.
\end{align}

Next, since $W_3 \in \{2,3\}$ almost surely, either $w_3=2$ or $w_3=3$.

If $w_3=2$, then $\{Y=2,W=w\}$ implies $X=3$, whereas under the uniform input distribution,
\begin{align}
    \Pr[X=1 \mid Y=2]=\Pr[X=3 \mid Y=2]=\frac12.
\end{align}
Again this contradicts $X-Y-W$.

If $w_3=3$, then $\{Y=3,W=w\}$ implies $X=3$, whereas
\begin{align}
    \Pr[X=2 \mid Y=3]=\Pr[X=3 \mid Y=3]=\frac12,
\end{align}
which also contradicts $X-Y-W$.

Hence no point in the support of $W$ can satisfy $w_1=1$. But this is impossible because $\Pr[W_1=1]=1/2$. Therefore, the channel in \eqref{singular bad matrix example appendix} is not perfectly representable.
\end{proof}

\begin{proof}[Proof of Lemma \ref{total variation lemma}]
By Lemma~\ref{optimality of canonical FR}, it is sufficient to maximize $I(Y;W)$ over all canonical functional representations
\begin{align}
    W=(W_1,W_2),
\end{align}
where
\begin{align}
    W_1 \sim p_{Y|X=1}, \qquad W_2 \sim p_{Y|X=2},
\end{align}
and $W$ is independent of $X$. Such a canonical functional representation is completely determined by a coupling of $W_1$ and $W_2$. Moreover, since $Y=W_X$, we have
\begin{align}
    I(Y;W)=H(Y)-H(Y|W).
\end{align}
Therefore, maximizing $I(Y;W)$ is equivalent to minimizing $H(Y|W)$ over all couplings of $W_1$ and $W_2$.

Now fix a coupling of $W_1$ and $W_2$. Then
\begin{align}
     & \ \ \ \ H(Y|W) \\
    &= H(W_X|W) \\
    &= \sum_{w_1,w_2} p_{W_1W_2}(w_1,w_2)\,
       H(W_X|W=(w_1,w_2)).
\end{align}
For each $(w_1,w_2)$,
\begin{align}
H(W_X|W=(w_1,w_2))
=
\begin{cases}
0, & \text{if } w_1=w_2,\\[0.3em]
H(X), & \text{if } w_1\neq w_2.
\end{cases}
\end{align}
Indeed, if $w_1=w_2$, then $W_X$ is deterministic given $W=(w_1,w_2)$; whereas if $w_1\neq w_2$, then $W_X$ is just a relabeling of $X$, and hence has entropy $H(X)$. Thus
\begin{align}
    H(Y|W)=\Pr[W_1\neq W_2]\,H(X).
\end{align}
Hence,
\begin{align}
    \min H(Y|W)
    = \Big(\min \Pr[W_1\neq W_2]\Big) H(X),
\end{align}
where the minimum is taken over all couplings of $W_1$ and $W_2$.

We now show that
\begin{align}
    \min \Pr[W_1\neq W_2]
    = \mathrm{TV}(p_{Y|X=1},p_{Y|X=2}) = r.
\end{align}
For any coupling,
\begin{align}
    \Pr[W_1=W_2]
    &= \sum_{y} \Pr[W_1=y,W_2=y] \\
    &\leq \sum_y \min\{p_{Y|X=1}(y),p_{Y|X=2}(y)\}.
\end{align}
Therefore,
\begin{align}
    \Pr[W_1\neq W_2]
    &\geq 1-\sum_y \min\{p_{Y|X=1}(y),p_{Y|X=2}(y)\} \\
    &= \mathrm{TV}(p_{Y|X=1},p_{Y|X=2}) \\
    &= r.
\end{align}
Moreover, equality is achieved by a maximal coupling. Hence
\begin{align}
    \min H(Y|W)=rH(X).
\end{align}
Consequently,
\begin{align}
    \max I(Y;Z)=\max I(Y;W)=H(Y)-rH(X).
\end{align}

For the second claim, recall that every functional representation satisfies
\begin{align}
    I(Y;Z)\leq H(Y|X)=H(Y)-I(X;Y).
\end{align}
Since we have shown that the maximum value of $I(Y;Z)$ equals $H(Y)-rH(X)$, it follows that
\begin{align}
    rH(X)\geq I(X;Y),
\end{align}
with equality if and only if the upper bound is attained, i.e., if and only if $p_{Y|X}$ is perfectly representable. Since $|\mathcal X|=2$, Lemma~\ref{singularity lemma} implies that this is equivalent to $p_{Y|X}$ being singular. Therefore,
\begin{align}
    rH(X)=I(X;Y)
\end{align}
if and only if $p_{Y|X}$ is singular.
\end{proof}

%% file: appendixB.tex
\begin{proof}[Proof of Lemma \ref{reordering operations lemma}]
Consider first a column $c\vec{p}$ with $0<c<1$ and $\vec{p}=(1,1)^T$. Let $\vec{e}_1=(1,0)^T$ and $\vec{e}_2=(0,1)^T$. Suppose that a VPS splits this column into $c\vec{e}_1$ and $c\vec{e}_2$, and a PPS then splits only $c\vec{e}_1$ in half. To obtain the same three columns in the desired order, first use PPS to split $c\vec{p}$ in half, then use VPS to separate the two rows of each copy. Finally, merge the two columns with pattern $\vec{e}_2$ by PPM. The two columns with pattern $\vec{e}_1$ remain separate, since they correspond to distinct output symbols. Thus, to decide which pieces to merge, we keep track of the output column to which each piece contributes.

Let a finite cascade take $M$ to $N$. Since the output alphabet of each channel is its output support, we work throughout with nonzero columns. Write the columns of $M$ as $c_i\vec{p}_i$, where $c_i>0$ is the value and $\vec{p}_i$ is the pattern of column $i$. For each intermediate matrix $Q$, we describe its columns using pieces of these original columns. For each $i$, take a finite family of copies indexed by $a$, with weights $w_{ia}>0$ such that $\sum_a w_{ia}=1$. For each copy, take nonzero patterns $\vec{p}_{iab}$ with pairwise disjoint supports such that $\sum_b\vec{p}_{iab}=\vec{p}_i$. The corresponding pieces are the columns $c_iw_{ia}\vec{p}_{iab}$. Denote the destination of such a piece by $\ell(i,a,b)$ and the pattern of column $j$ of $Q$ by $\vec{q}_j$. We require that
\begin{align*}
    \vec{p}_{iab} &= \vec{q}_{\ell(i,a,b)},\\
    Q[:,j] &= \sum_{\ell(i,a,b)=j} c_iw_{ia}\vec{p}_{iab}.
\end{align*}
Thus, each piece has the pattern of its destination, and the pieces assigned to each column sum to that column. We also record the proportional splits that produce the weights and the support splits that produce the pattern partition of each copy.

We show by induction along the given cascade that such a description exists. Initially $Q=M$. For each $i$, take one copy of weight one with the single pattern $\vec{p}_i$ and destination $i$. Now suppose the description holds for $Q$ and consider the next operation.

\begin{itemize}
    \item If a PPS splits column $l$ in the ratio $\alpha:(1-\alpha)$, replace each weighted copy of every original column by two copies of weights $\alpha w_{ia}$ and $(1-\alpha)w_{ia}$. Give both copies the same support partition as before. For a piece whose destination was $l$, send the two new pieces to the two new columns, respectively. For every other piece, send both new pieces to its original destination. Their contributions there add to the original contribution, while the pieces previously assigned to $l$ give the required proportions in the two new columns.
    \item If a VPS splits the pattern of column $l$ into two nonzero patterns with disjoint supports, every piece assigned to $l$ has that same pattern $\vec{q}_l$. Split each such piece according to the same two supports, keeping its value and assigning its two parts to the corresponding new columns. This refines the support partition of each affected copy and gives the required sums at both destinations.
    \item If a PPM merges two columns, simply assign the pieces of both columns to their common destination. The patterns agree by the definition of PPM, and the pieces sum to the merged column.
\end{itemize}
This completes the induction.

We now realize the description of $N$ in the required order. Starting from each column $c_i\vec{p}_i$ of $M$, perform the recorded proportional splits to obtain the columns $c_iw_{ia}\vec{p}_i$. Next, perform the recorded support splits on each copy to obtain $c_iw_{ia}\vec{p}_{iab}$. Each such split has two nonempty supports, and the value of the copy is at most the value of the column on which the original VPS was performed. Finally, for each column $j$ of $N$, merge the pieces with destination $j$ by PPM. They all have pattern $\vec{q}_j$, and their sum is $N[:,j]$. Since each output column is nonzero, it receives at least one piece. This gives $N$ up to a permutation of its columns. The given cascade is finite and each operation records only finitely many splits or reassignments, so the resulting cascade of PPS, followed by VPS and then PPM, is finite.
\end{proof}

\begin{proof}[Proof of Lemma \ref{preserving operations lemma}]
Let $M$ be perfectly representable, and let
\begin{align}
    W=(W_1,\ldots,W_k)
\end{align}
be a perfect canonical functional representation of the corresponding pair $(X,Y)$. Thus,
\begin{align}
    I(X;W)=0, \qquad Y=W_X, \qquad X-Y-W,
\end{align}
and
\begin{align}
    W_x \sim p_{Y|X}(\cdot|x), \qquad x\in\mathcal X.
\end{align}
We show separately that each of the three operations preserves perfect representability.

\medskip

\noindent\textbf{1) PPS preserves perfect representability.}
Let $A_{l,\alpha}$ split the symbol $l\in\mathcal Y$ into two new symbols $l_1,l_2$, with probabilities\vadjust{\vskip 2.2pt} $\alpha$ and $1-\alpha$, respectively. Let $B\sim\mathrm{Bern}(\alpha)$ be independent of $(X,Y,W)$, and define
\begin{align}
    \phi_B(y)
    \triangleq
    \begin{cases}
        y, & y\notin\{l\},\\
        l_1, & y=l,\; B=1,\\
        l_2, & y=l,\; B=0.
    \end{cases}
\end{align}
Now define
\begin{align}
    Y' \triangleq \phi_B(Y), \qquad W'_x \triangleq \phi_B(W_x), \qquad x\in\mathcal X.
\end{align}
Then $W'$ is independent of $X$ since it is a function of $(W,B)$, which is independent of $X$, and
\begin{align}
    W'_X=\phi_B(W_X)=\phi_B(Y)=Y'.
\end{align}
Moreover, for each $x$,
\begin{align}
    W'_x \sim p_{Y'|X}(\cdot|x),
\end{align}
where $p_{Y'|X}$ is exactly the channel obtained from $p_{Y|X}$ by applying $A_{l,\alpha}$.

It remains to show that $X-Y'-W'$. If $y'\notin\{l_1,l_2\}$, then $Y'=y'$ is equivalent to $Y=y'$, so
\begin{align}
    p_{W'|X,Y'}(\cdot|x,y')
    = p_{W'|X,Y}(\cdot|x,y')
    = p_{W'|Y}(\cdot|y'),
\end{align}
where the last equality follows from $X-Y-W$ and the independence of $B$.
If $y'=l_1$, then $Y'=l_1$ is equivalent to $\{Y=l,B=1\}$, and hence
\begin{align}
    p_{W'|X,Y'}(\cdot|x,l_1)
    = p_{W'|X,Y,B}(\cdot|x,l,1)
    = p_{W'|Y,B}(\cdot|l,1),
\end{align}
which does not depend on $x$. The case $y'=l_2$ is identical. Thus $X-Y'-W'$, so $A_{l,\alpha}(M)$ is perfectly representable.

\medskip

\noindent\textbf{2) PPM preserves perfect representability.}
Let $B_{l_1,l_2}$ merge two symbols $l_1,l_2\in\mathcal Y$ that have the same pattern. Since $M$ is perfectly representable, it is singular by Lemma~\ref{singularity lemma}. Hence the columns corresponding to $l_1$ and $l_2$ have values $\alpha,\beta>0$ and the same support set
\begin{align}
    S \triangleq \{x\in\mathcal X : p_{Y|X}(l_1|x)>0\} \\
    = \{x\in\mathcal X : p_{Y|X}(l_2|x)>0\}
\end{align}
with
\begin{align}
    p_{Y|X}(l_1|x)=\alpha,\qquad p_{Y|X}(l_2|x)=\beta,\qquad x\in S.
\end{align}
Let $l$ be the merged symbol, and define
\begin{align}
    \psi(y)
    \triangleq
    \begin{cases}
        y, & y\notin\{l_1,l_2\},\\
        l, & y\in\{l_1,l_2\}.
    \end{cases}
\end{align}
Set
\begin{align}
    Y' \triangleq \psi(Y), \qquad W'_x \triangleq \psi(W_x), \qquad x\in\mathcal X.
\end{align}
Then $W'$ is independent of $X$, and
\begin{align}
    W'_X=\psi(W_X)=\psi(Y)=Y'.
\end{align}
Also, for each $x$,
\begin{align}
    W'_x \sim p_{Y'|X}(\cdot|x),
\end{align}
where $p_{Y'|X}$ is the merged channel.

It remains to show that $X-Y'-W'$. For $y'\neq l$, the argument is the same as in the PPS case. Now consider $y'=l$. If $x\notin S$, then $\Pr[Y'=l|X=x]=0$, so there is nothing to prove. If $x\in S$, then
\begin{align}
    \Pr[Y=l_1|X=x,Y'=l] = \frac{\alpha}{\alpha+\beta},\\
    \Pr[Y=l_2|X=x,Y'=l] = \frac{\beta}{\alpha+\beta},
\end{align}
which do not depend on $x$. Therefore, for any measurable set $\mathcal A$,
\begin{align}
    &\Pr[W'\in\mathcal A|X=x,Y'=l]\\
    &= \frac{\alpha}{\alpha+\beta}\Pr[W'\in\mathcal A|X=x,Y=l_1] \nonumber\\
    &\quad + \frac{\beta}{\alpha+\beta}\Pr[W'\in\mathcal A|X=x,Y=l_2].
\end{align}
Since $W'$ is a function of $W$ and $X-Y-W$,
\begin{align}
    \Pr[W'\in\mathcal A|X=x,Y=l_j]
    = \Pr[W'\in\mathcal A|Y=l_j]
\end{align}
for $j \in \{1,2\}$. Hence $\Pr[W'\in\mathcal A|X=x,Y'=l]$ does not depend on $x$, and thus $X-Y'-W'$. Therefore, $B_{l_1,l_2}(M)$ is perfectly representable.

\medskip

\noindent\textbf{3) VPS preserves perfect representability.}
Let $C_{l,\mathbf b}$ split the symbol $l\in\mathcal Y$ into two new symbols $l_1,l_2$ according to a fixed vector
\begin{align}
    \mathbf b=(b_1,\ldots,b_k)\in\{0,1\}^k.
\end{align}
For each $x\in\mathcal X$, define
\begin{align}
    \eta_x(y)
    \triangleq
    \begin{cases}
        y, & y\neq l,\\
        l_1, & y=l,\; b_x=1,\\
        l_2, & y=l,\; b_x=0.
    \end{cases}
\end{align}
Set
\begin{align}
    W'_x \triangleq \eta_x(W_x), \qquad x\in\mathcal X,
\end{align}
and define
\begin{align}
    Y' \triangleq W'_X = \eta_X(Y).
\end{align}
Since $W'$ is a deterministic function of $W$, it is independent of $X$. Also, $W'_X=Y'$ by construction, and for each $x$,
\begin{align}
    W'_x \sim p_{Y'|X}(\cdot|x),
\end{align}
where $p_{Y'|X}$ is exactly the channel obtained from $p_{Y|X}$ by applying $C_{l,\mathbf b}$.

It remains to show that $X-Y'-W'$. Let
\begin{align}
    S_1 \triangleq \{x\in\mathcal X : b_x=1\}, \qquad S_2 \triangleq \{x\in\mathcal X : b_x=0\}.
\end{align}
If $y'\notin\{l_1,l_2\}$, then again $Y'=y'$ is equivalent to $Y=y'$, and the claim follows from $X-Y-W$. If $y'=l_1$, then $Y'=l_1$ is equivalent to $\{Y=l,\; X\in S_1\}$. Since $X-Y-W$, the random variable $W$ is independent of $X$ given $Y=l$, and therefore any function of $W$, in particular $W'$, is also independent of $X$ given $Y=l$. Hence, for every $x\in S_1$,
\begin{align}
    p_{W'|X,Y'}(\cdot|x,l_1)
    = p_{W'|X,Y}(\cdot|x,l)
    = p_{W'|Y}(\cdot|l),
\end{align}
which does not depend on $x$. The case $y'=l_2$ is identical. Thus $X-Y'-W'$, and so $C_{l,\mathbf b}(M)$ is perfectly representable.

Since PPS, PPM, and VPS each preserve perfect representability, the lemma follows.
\end{proof}

\begin{proof}[Proof of Lemma \ref{preserving operations lemma not}]
This is an immediate consequence of Lemma~\ref{preserving operations lemma} and the fact that PPS and PPM are inverses of each other, while VPS and VPM are inverses of each other.

We prove the claim by contraposition. Suppose that applying one of the operations PPM, PPS, or VPM to a channel $M$ yields a perfectly representable channel $M'$. Since the inverse of that operation is, respectively, PPS, PPM, or VPS, Lemma~\ref{preserving operations lemma} implies that applying the inverse operation to $M'$ also yields a perfectly representable channel. But applying the inverse operation to $M'$ recovers $M$. Hence $M$ must be perfectly representable.

Therefore, if $M$ is not perfectly representable, then the result of applying PPM, PPS, or VPM to $M$ cannot be perfectly representable.
\end{proof}

\begin{proof}[Proof of Theorem~\ref{main theorem}]

Recall that $\mathbf{1}_k$ is the all ones vector of size $k$ and that $\mathbf{1}_k$ is perfectly representable. Thus, the ``if'' part of the proof is a direct result of Lemma \ref{preserving operations lemma} and Lemma \ref{preserving operations lemma not} by applying the inverse of the operations mentioned in Lemma \ref{preserving operations lemma} to reduce $M$ to $\mathbf{1}_k$. Example \ref{reducing example} illustrates this reduction.

    For the ``only if'' part, we show that if a matrix $M$ associated with $p_{Y|X}$ is perfectly representable, then $\mathbf{1}_k$ could be obtained from $M$ with repeated use of PPM, VPM, and PPS operations. Since $M$ is perfectly representable, there exists a perfect canonical functional representation $W$.   Without loss of generality, we may assume that for each $w$, $p_W(w)>0$.

   First, let $Y' = (Y,W)$, so we have $Y-Y'-W$ and
   \begin{align}
       p_{W|Y'}(w'|(y,w)) = \mathds{1}\{w'=w\},\\
       p_{Y'|Y}((y,w)|y) = p_{W|Y}(w|y).
   \end{align}
Note that $W$ is a perfect functional representation for $(X,Y')$ since $X - Y' - W$ and
\begin{align}
    Y'=(Y,W) = (g(X,W),W) =: g'(X,W).
\end{align}
Moreover, $Y$ is a deterministic function of $Y'$ and the matrix $p_{Y'|X}$ can be obtained from the matrix $p_{Y|X}$ by the PPS operation, so it is sufficient to show that $\mathbf{1}_k$ can be obtained from $p_{Y'|X}$ via PPM and VPM operations.  As PPS is the inverse of PPM, and VPS is the inverse of VPM, it thus suffices to show that the matrix $p_{Y'|X}$ can be obtained from $\mathbf1_k$ via PPS and VPS operations.  This is shown in the following paragraphs.

Fix $w \in \mathcal{W}$ and let
\begin{align}
    \mathcal{Y}_w' := \{ (y,w) \in \mathcal{Y}' : y\in\mathcal Y \}.
\end{align}
Note that $\mathcal{Y'}$ can be partitioned into the sets $\mathcal{Y}_w'$. Since $W$ is a functional representation of $(X,Y)$, for a fixed $x$ and fixed $w$, there is a unique $(y,w) \in \mathcal{Y}_w'$, namely $(y,w)=g'(x,w)$, such that
\begin{align}
    p_{Y'|X}((y,w)|x) > 0,
\end{align}
and, for any $\tilde y \neq y$,
\begin{align}
    p_{Y'|X}((\tilde y,w)|x) =0.
\end{align}
Moreover,
\begin{align}
    p_{Y'|X}(g'(x,w)|x) = p_W(w)>0.
\end{align}
Hence, $p_{Y'|X}(g'(x,w)|x)$ does not depend on $x$.

Finally, for each $y' \in \mathcal{Y}_w'$, let $S_{y'}$ be the set
\begin{align}
    S_{y'} = \{ x \in \mathcal{X} : g'(x,w) = y' \}.
\end{align}
Since the set
\begin{align}
    \{p_{Y'|X}(y'|x) \cdot p_{W|Y'}(w|y') : y' \in \mathcal{Y}' \}
\end{align}
has exactly one non-zero element for any $(x,w)$ pair, for each $w$, the collection $\{S_{y'} : y' \in \mathcal{Y}_w' \}$ is a partition of $\mathcal{X}$. For each $y' \in \mathcal{Y}_w'$, let $\vec{b}_{y'}$ be the indicator vector of $S_{y'}$, with entries $\vec{b}_{y'}(x)=\mathds{1}\{x\in S_{y'}\}$ for $x\in\mathcal{X}$. %
We can apply PPS operation to $\mathbf{1}_k$ with coefficients $c_w=p_W(w)$, and get vectors with patterns $\mathbf{1}_k$ and values $c_w$ for all $w \in \mathcal{W}$. Then, for all $w \in \mathcal{W}$, we can apply VPS operation to the column with pattern $\mathbf{1}_k$ and value $c_w$ with coefficients $\vec{b}_{y'}$ and get $p_{Y'|X}$. Figure~\ref{figure for PPS,VPS} illustrates these operations in reverse.
\begin{figure}[h]
       \centering
       \includegraphics[width=\linewidth]{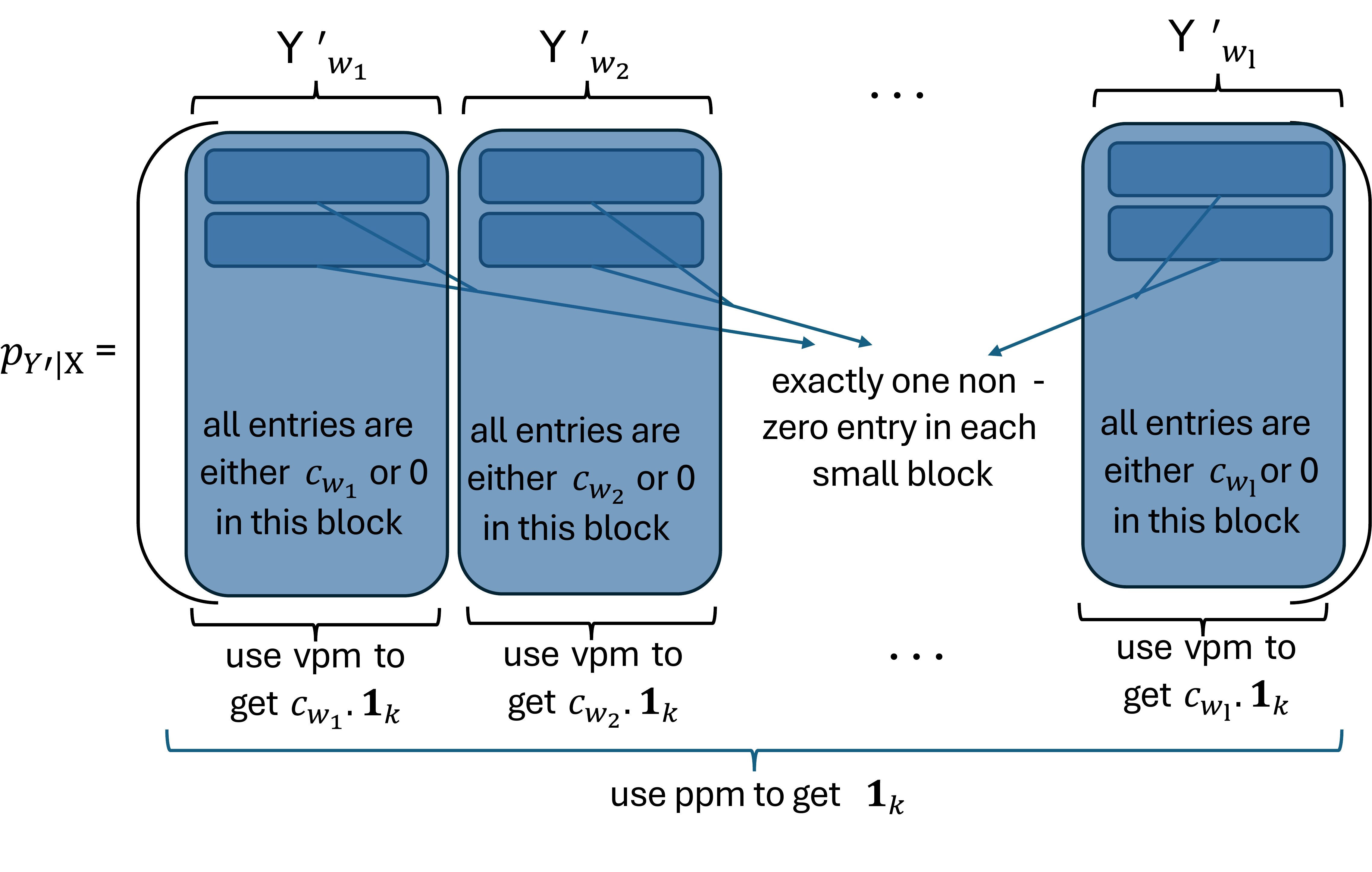}
       \caption{Using VPM and PPM to obtain $\mathbf{1}_k$ from $p_{Y'|X}$}
       \label{figure for PPS,VPS}
\end{figure}
\end{proof}

%% file: appendixC.tex
\begin{proof}[Proof of Theorem \ref{LP algorithm theorem}]
In Part~I, we show that $M$ is perfectly representable if and only if the extended system ($\widehat{H}\widehat{\vec{x}}=\widehat{\vec{v}}$) has a nonnegative solution. In Part~II, we show that this is equivalent to the existence of a nonnegative solution to $H\vec{x}=\vec{v}$. In Part~III, we show that such a solution exists if and only if the linear program in Theorem~\ref{LP algorithm theorem} has optimal value one.

\medskip
\noindent\textit{Part I. Perfect representability and feasibility of the extended system.}

By Theorem~\ref{main theorem}, $M$ is perfectly representable if and only if it can be obtained from $\mathbf{1}_k$ by PPS, VPS, and PPM. First, suppose that $M$ is perfectly representable. By Lemma~\ref{reordering operations lemma}, we can perform all PPS operations first, followed by the VPS operations and finally the PPM operations.

Each column obtained after the PPS operations has pattern $\mathbf{1}_k$. If such a column has value $\beta$, the columns obtained from it by VPS all have value $\beta$, and their patterns have disjoint supports and add up to $\mathbf{1}_k$. Thus, these patterns form one of the partitions $\widehat{S}_1,\ldots,\widehat{S}_{B(k)}$. For each $j$, let $\widehat{x}_j$ be the sum of the values of the PPS columns that lead to $\widehat{S}_j$, with $\widehat{x}_j=0$ if there are no such columns. When we merge columns with the same pattern by PPM, each partition containing $\widehat{\vec{p}}_i$ contributes its weight to that pattern. Hence,
\begin{align}
    \sum_{j:\,\widehat{\vec{p}}_i\in\widehat{S}_j}\widehat{x}_j
    =\widehat{v}_i,\qquad i\in\{1,\ldots,2^k-1\},
    \label{full partition system}
\end{align}
where the sum is zero for any pattern that does not occur in $M$. This gives a nonnegative solution to~\eqref{full partition system}.

Conversely, suppose that $\widehat{\vec{x}}\geq0$ satisfies~\eqref{full partition system}. Fix a row $\ell\in[k]$ and sum the equations over all patterns whose $\ell$-th entry is one. Since each partition contains exactly one such pattern, the left-hand side is $\sum_{j=1}^{B(k)}\widehat{x}_j$. The right-hand side is the sum of row $\ell$ of $M$, which is one. We can therefore obtain $M$ from $\mathbf{1}_k$ as follows:
\begin{enumerate}
    \item Starting from $\mathbf{1}_k$, use PPS to obtain one column of value $\widehat{x}_j$ for each $j$ with $\widehat{x}_j>0$, all with pattern $\mathbf{1}_k$.
    \item For each such $j$, use VPS to split that column according to the partition $\widehat{S}_j$.
    \item Finally, merge all columns with the same pattern using PPM.
\end{enumerate}
By~\eqref{full partition system}, the resulting column for each pattern occurring in $M$ has its prescribed value, and no other pattern occurs. This recovers $M$ up to the order of its columns, so $M$ is perfectly representable. We now use the zero entries of $\widehat{\vec{v}}$ to see which partition weights can be omitted.

\medskip
\noindent\textit{Part II. Equivalence of the extended and reduced systems.}

The full system in~\eqref{full partition system} is feasible if and only if there exists
\begin{align}
    \vec{x}\geq0
\end{align}
such that
\begin{align}
    H\vec{x}=\vec{v}.
    \label{sparse partition system}
\end{align}
Indeed, suppose that $\widehat{H}\widehat{\vec{x}}
=\widehat{\vec{v}}$. For every $i>m$, we have
\begin{align}
    0
    =
    \widehat{v}_i
    =
    \sum_{j=1}^{B(k)}
    \widehat{H}_{ij}\widehat{x}_j.
\end{align}
Since every term in this sum is nonnegative, any partition containing a pattern that does not occur in $M$ must have zero weight. Therefore, only the columns indexed by $\mathcal{J}$ can receive positive weight, and restricting to these columns gives~\eqref{sparse partition system}. Conversely, any solution of~\eqref{sparse partition system} can be extended by assigning zero weight to each partition containing a pattern that does not occur in $M$, giving a solution of~\eqref{full partition system}.

Moreover, for the partitions indexed by $\mathcal{J}$, any vector of weights $\vec{x}\geq 0$ satisfying
\begin{align}
    H\vec{x}=\vec{v}
\end{align}
automatically satisfies
\begin{align}
    \sum_{j=1}^{R}x_j=1.
\end{align}
Indeed, fix any row $\ell\in[k]$ and sum the equations in
$H\vec{x}=\vec{v}$ over all occurring patterns containing
$\ell$. Since every $S_j$ contains exactly one pattern whose
$\ell$-th entry is one, the left-hand side equals
$\sum_{j=1}^{R}x_j$, while the right-hand side equals one by
row stochasticity.

\medskip
\noindent\textit{Part III. Equivalence to the linear program.}

Let us define $P$ as
\begin{align}
    P \triangleq [\vec{p}_1 \ \vec{p}_2 \ \cdots \ \vec{p}_m],
\end{align}
so that $P$ is a $k\times m$ matrix whose columns are the distinct nonzero binary patterns appearing in $M$.
By construction, each set $S_j$ is a partition of $\mathbf{1}_k$, and therefore, for every $j\in\{1,\dots,R\}$,
\begin{align}
    \sum_{i=1}^m H_{ij}\vec{p}_i=\mathbf{1}_k.
\end{align}
Equivalently,
\begin{align}
    PH=[\mathbf{1}_k \ \mathbf{1}_k \ \cdots \ \mathbf{1}_k],
\end{align}
a $k\times R$ matrix whose every column is $\mathbf{1}_k$.

Now let $\vec{x}\geq 0$ satisfy $H\vec{x}\leq \vec{v}$. Multiplying by $P$, we get
\begin{align}
    P H\vec{x}\leq P\vec{v}. \label{eq:LP-thm-ineq1}
\end{align}
Since the rows of the conditional probability matrix $M$ sum to $1$, we have
\begin{align}
    P\vec{v}=\mathbf{1}_k.
\end{align}
Also, using the form of $PH$ above,
\begin{align}
    PH\vec{x}
    = (\vec{1}_R\cdot \vec{x})\mathbf{1}_k.
\end{align}
Substituting into \eqref{eq:LP-thm-ineq1}, we obtain
\begin{align}
    (\vec{1}_R\cdot \vec{x})\mathbf{1}_k \leq \mathbf{1}_k,
\end{align}
and hence
\begin{align}
    \vec{1}_R\cdot \vec{x}\leq 1.
\end{align}
Therefore,
\begin{align}
    \max_{\substack{\vec{x}\geq 0\\ H\vec{x}\leq \vec{v}}}\vec{1}_R\cdot \vec{x}\leq 1.
\end{align}

If there exists $\vec{x}\geq 0$ such that $H\vec{x}=\vec{v}$, then this $\vec{x}$ is feasible and satisfies
\begin{align}
    (\vec{1}_R\cdot \vec{x})\mathbf{1}_k
    = PH\vec{x}
    = P\vec{v}
    = \mathbf{1}_k,
\end{align}
so $\vec{1}_R\cdot \vec{x}=1$. Hence,
\begin{align}
    \max_{\substack{\vec{x}\geq 0\\ H\vec{x}\leq \vec{v}}}\vec{1}_R\cdot \vec{x}=1.
\end{align}

Conversely, suppose
\begin{align}
    \max_{\substack{\vec{x}\geq 0\\ H\vec{x}\leq \vec{v}}}\vec{1}_R\cdot \vec{x}=1,
\end{align}
and let $\vec{x}^*$ be an optimal solution. Then
\begin{align}
    PH\vec{x}^* = (\vec{1}_R\cdot \vec{x}^*)\mathbf{1}_k = \mathbf{1}_k = P\vec{v}.
\end{align}
Thus,
\begin{align}
    P(\vec{v}-H\vec{x}^*)=0.
\end{align}
Since $\vec{v}-H\vec{x}^*\geq 0$ and every column $\vec{p}_i$ of $P$ is a nonzero binary vector, the only nonnegative vector in the kernel of $P$ is the zero vector. Therefore,
\begin{align}
    \vec{v}-H\vec{x}^*=0,
\end{align}
that is,
\begin{align}
    H\vec{x}^*=\vec{v}.
\end{align}
Hence there exists $\vec{x}\geq 0$ such that $H\vec{x}=\vec{v}$.
\end{proof}

%% file: appendixD.tex
\subsection{Complexity of Solving the Dual LP}
\label{dual complexity proof subsection}

\begin{proof}[Proof of Theorem~\ref{dual complexity theorem}]
Identify each pattern $\vec{p}_i$ with its support
\begin{align}
    A_i:=\{\ell\in[k]:\vec{p}_i(\ell)=1\},
\end{align}
and let $\mathcal A:=\{A_1,\ldots,A_m\}$.
Write $\Pi(\mathcal A)$ for the collection of all set
partitions of $[k]$ whose blocks belong to $\mathcal A$.
Each $\pi\in\Pi(\mathcal A)$ gives the dual constraint
\begin{align}
    \sum_{A_i\in\pi}y_i\geq1.
\end{align}

A separation oracle either certifies that a candidate vector
is feasible or returns a violated constraint. Nonnegativity
is checked directly. For $\vec{y}\geq0$, define
\begin{align}
    \gamma(\vec{y})
    :=
    \min_{\pi\in\Pi(\mathcal A)}
    \sum_{A_i\in\pi}y_i,
    \label{minimum partition weight}
\end{align}
where an empty minimum equals $+\infty$.
If $\gamma(\vec{y})<1$, a minimizing partition gives a
violated constraint. Otherwise, all partition constraints
are satisfied.

We now focus on a constraint-generation implementation.
Initialize $\Pi_0=\emptyset$, so that the first restricted
LP contains only the constraints $\vec{y}\geq0$.
After $t$ partition constraints have been added, perform the
following steps:
\begin{enumerate}
    \item Solve the restricted LP
    \begin{align}
        \min_{\vec{y}\geq0}\quad
        &\vec{v}\cdot\vec{y}
        \nonumber\\
        \text{subject to}\quad
        &\sum_{A_i\in\pi}y_i\geq1,
        \qquad \pi\in\Pi_t,
    \end{align}
    and let $\vec{y}^{(t)}$ be an optimal solution.

    \item Compute $\gamma(\vec{y}^{(t)})$.
    If $\gamma(\vec{y}^{(t)})\geq1$, stop and return
    $\vec{y}^{(t)}$.

    \item Otherwise, choose a minimizing partition
    $\pi_t^\star$, add its constraint by setting
    \begin{align}
        \Pi_{t+1}:=\Pi_t\cup\{\pi_t^\star\},
    \end{align}
    and repeat.
\end{enumerate}

Since every occurring-pattern value $v_i$ is positive, the
initial optimum is $\vec{y}^{(0)}=0$. At termination, the
returned vector is feasible for the full dual LP and optimal
for a relaxation of that LP, so it is also optimal for the
full LP. Each added constraint is new, because the current
solution satisfies all previously added constraints.
Consequently, this implementation terminates after at most
$R$ additions.

We next describe how to compute the separation oracle.
For every $T\subseteq[k]$, let $D(T)$ be the minimum cost
of partitioning $T$ using sets from $\mathcal A$. Set
\begin{align}
    D(\emptyset)=0.
\end{align}
For nonempty $T$, let $a(T)$ denote its smallest element and
set
\begin{align}
    D(T)
    =
    \min_{\substack{i:\ A_i\subseteq T\\
    a(T)\in A_i}}
    \left\{y_i+D(T\setminus A_i)\right\}.
    \label{subset DP recurrence}
\end{align}
An empty minimum is interpreted as $+\infty$.

Every partition of $T$ contains exactly one block containing
$a(T)$. The recurrence considers all possible choices for
this block and recursively partitions the remaining elements.
Evaluating states in increasing order of their cardinality
therefore gives
\begin{align}
    D([k])=\gamma(\vec{y}).
\end{align}
Storing a minimizing choice at each finite state also allows
a minimizing partition to be recovered.

Representing subsets as bitmasks, there are $2^k$ states.
At each state, scan at most $m$ patterns and retain the
smallest admissible candidate value. No sorting is needed.
Thus, one separation call requires
\begin{align}
    2^k O(m)
\end{align}
arithmetic and bitmask operations.

Finally, we analyze the total computational complexity.
Let $T_{\mathrm{LP}}(m,m+t,L)$ denote the cost of solving
the restricted LP with $m$ variables and $m+t$ constraints,
including nonnegativity. Polynomial-time algorithms for
rational linear programming give
\begin{align}
    T_{\mathrm{LP}}(m,m+t,L)
    =\operatorname{poly}(m+t,L)
\end{align}
for these explicitly represented LPs~\cite{groetschellovaszschrijver1988}.
If the constraint-generation implementation performs $N_I$
solve-and-separate iterations, including its final feasibility
check, its arithmetic-operation cost is bounded by
\begin{align}
    \sum_{t=0}^{N_I-1}
    \left[
        T_{\mathrm{LP}}(m,m+t,L)+2^k O(m)
    \right].
\end{align}

For a worst-case guarantee, use the same oracle within the
ellipsoid method: the optimization--separation results for
rational polyhedra give exact optimization using
$\operatorname{poly}(m,L)$ separation calls and polynomial
additional computation~\cite{groetschellovaszschrijver1981,groetschellovaszschrijver1988}. Note that this call bound applies to the ellipsoid-based algorithm,
not to the restricted-LP loop above.

Combining this guarantee with the oracle cost, and absorbing
exact rational-arithmetic costs into the polynomial factors,
the total running time of the ellipsoid-based algorithm is
\begin{align}
    T_{\mathrm{total}}
    &\leq 2^k O(m)\operatorname{poly}(m,L)
    \nonumber\\
    &\quad+\operatorname{poly}(m,L)
    \nonumber\\
    &=2^k\operatorname{poly}(m,L).
\end{align}
\end{proof}

\subsection{NP-Hardness of Separation and Exact LP Optimization}
\label{RXC3 appendix subsection}

We now justify the claim in Section~\ref{dual separation subsection} that exact optimization of the compactly represented LP is NP-hard when $k$ is part of the input.

Restricted Exact Cover by 3-Sets (RXC3) is the following decision problem. The input consists of a ground set
\begin{align}
    U=[k],
    \qquad
    k=3q,
\end{align}
and a collection
\begin{align}
    \mathcal C
    \subseteq
    \binom{[k]}{3}
\end{align}
such that every element of $[k]$ belongs to exactly three members of $\mathcal C$. The question is whether there exists a subcollection of pairwise disjoint triples whose union is $[k]$. RXC3 is NP-complete~\cite[Theorem A.1]{Gonzalez1985}.

Given an RXC3 instance, use the triples themselves as the occurring patterns and assign value $1/3$ to every pattern:
\begin{align}
    \mathcal A=\mathcal C,
    \qquad
    v_C=\frac{1}{3},
    \quad C\in\mathcal C.
\end{align}
Equivalently, define the corresponding matrix by
\begin{align}
    M_{\ell,C}
    =
    \frac{1}{3}
    \mathds{1}\{\ell\in C\}.
    \label{RXC3 matrix}
\end{align}
Since every element $\ell\in[k]$ belongs to exactly three triples, every row of $M$ sums to one. Moreover, every positive entry in a column equals $1/3$, so $M$ is a valid singular conditional probability matrix.

Consider the separation problem for the reduced dual LP at
\begin{align}
    \vec{y}=0.
\end{align}
A violated partition constraint exists if and only if there is a partition $\pi\in\Pi(\mathcal A)$ satisfying
\begin{align}
    \sum_{C\in\pi}y_C<1.
\end{align}
At $\vec{y}=0$, this inequality holds for every exact partition. Therefore,
\begin{align}
    \text{a violated constraint exists at }\vec{y}=0
    \quad\Longleftrightarrow\quad
    \Pi(\mathcal A)\neq\emptyset.
\end{align}
Since the occurring patterns are exactly the triples in $\mathcal C$, the condition $\Pi(\mathcal A)\neq\emptyset$ holds if and only if $\mathcal C$ contains an exact cover of $[k]$. Thus, deciding whether a violated dual constraint exists is NP-hard. It is, in fact, NP-complete, because an exact partition is a polynomial-size certificate.

This hardness also appears in the natural constraint-generation implementation initialized with no partition constraints. Its initial restricted dual is
\begin{align}
    \min_{\vec{y}\geq0}
    \frac{1}{3}
    \sum_{C\in\mathcal C}y_C.
\end{align}
Because every objective coefficient is positive, its unique optimum is
\begin{align}
    \vec{y}^{(0)}=0.
\end{align}
Consequently, the first separation call of this implementation determines whether the RXC3 instance has an exact cover.

The same construction proves that exact optimization of the compactly represented LP is NP-hard. Suppose first that the RXC3 instance has no exact cover. Then
\begin{align}
    \Pi(\mathcal A)=\emptyset.
\end{align}
The reduced dual has no partition constraints, so $\vec{y}=0$ is feasible and its optimal value is zero.

Suppose instead that an exact cover
\begin{align}
    \pi\in\Pi(\mathcal A)
\end{align}
exists. Every dual-feasible vector $\vec{y}$ must satisfy
\begin{align}
    \sum_{C\in\pi}y_C\geq1.
\end{align}
Since $\vec{y}\geq0$,
\begin{align}
    \vec{v}\cdot\vec{y}
    &=
    \frac{1}{3}
    \sum_{C\in\mathcal C}y_C\\
    &\geq
    \frac{1}{3}
    \sum_{C\in\pi}y_C\\
    &\geq
    \frac{1}{3}.
\end{align}
Therefore, an exact optimizer would distinguish instances with optimal value zero from instances with optimal value at least $1/3$, and would solve RXC3. Exact optimization of the compact dual is consequently NP-hard. By strong duality, the same conclusion applies to exact optimization of the primal LP.

This result rules out a polynomial-time exact algorithm for variable $k$ unless $\mathrm{P}=\mathrm{NP}$. It does not show that the running time
\begin{align}
    2^k\operatorname{poly}(m,L)
\end{align}
from Section~\ref{dual separation subsection} is optimal. In particular, NP-hardness alone does not exclude a smaller exponential dependence or a subexponential parameterized algorithm.

This RXC3 reduction also does not by itself settle the complexity of deciding perfect representability. It distinguishes optimal value zero from optimal value at least $1/3$, whereas perfect representability asks whether the optimal value equals one. The separate reduction in Appendix~\ref{decision complexity proof subsection} addresses this decision problem.

\subsection{NP-Completeness of Deciding Perfect Representability}
\label{decision complexity proof subsection}

\begin{proof}[Proof of Theorem~\ref{perfect representability NP-hard theorem}]
We first give a polynomial-time reduction from 3-Colorability, which is NP-complete~\cite{garey_johnson_1979}.

Let
\begin{align}
    G=(V,E)
\end{align}
be a finite simple graph. Isolated vertices may be deleted without changing whether $G$ is 3-colorable. If no edges remain, replace $G$ by the graph consisting of a single edge, which is also 3-colorable. Hence, assume that $G$ has at least one edge and no isolated vertices.

Let
\begin{align}
    \mathcal C:=\{0,1,2\}
\end{align}
be the color set. Define the row set
\begin{align}
    \Omega_G
    :=
    \{a_u:u\in V\}
    \cup
    \{b_{e,c}:e\in E,\ c\in\mathcal C\}.
\end{align}

For every vertex $u\in V$ and color $c\in\mathcal C$, define the vertex-choice pattern
\begin{align}
    A_{u,c}
    :=
    \{a_u\}
    \cup
    \{b_{e,c}:e\in E
    \text{ is incident to }u\}.
    \label{vertex choice pattern}
\end{align}
For every edge $e\in E$ and color $c\in\mathcal C$, define the filler pattern
\begin{align}
    F_{e,c}:=\{b_{e,c}\}.
    \label{filler pattern}
\end{align}
Let
\begin{align}
    \mathcal A_G
    :=
    \{A_{u,c}:u\in V,\ c\in\mathcal C\}
    \cup
    \{F_{e,c}:e\in E,\ c\in\mathcal C\}.
\end{align}
Assign value $1/3$ to every pattern. Equivalently, define the matrix $M(G)$ by
\begin{align}
    M(G)_{\omega,B}
    :=
    \frac{1}{3}
    \mathds{1}\{\omega\in B\},
    \qquad
    \omega\in\Omega_G,\quad B\in\mathcal A_G.
    \label{graph matrix construction}
\end{align}

Every row belongs to exactly three patterns. The row $a_u$ belongs to
\begin{align}
    A_{u,0},\quad A_{u,1},\quad A_{u,2},
\end{align}
while, for an edge $e=\{u,w\}$, the row $b_{e,c}$ belongs to
\begin{align}
    A_{u,c},\quad A_{w,c},\quad F_{e,c}.
\end{align}
Thus, every row of $M(G)$ sums to one. Moreover, every positive entry in a column equals $1/3$, so $M(G)$ is singular.

The patterns in $\mathcal A_G$ are distinct. Patterns associated with different vertices are distinguished by their private rows $a_u$. For a fixed vertex $u$, the three patterns $A_{u,0},A_{u,1},A_{u,2}$ are distinguished by the color-indexed rows corresponding to any incident edge. The filler patterns are distinct singletons and contain no private vertex row.

We next show that the admissible exact partitions formed from $\mathcal A_G$ correspond exactly to proper 3-colorings of $G$.

Let
\begin{align}
    \pi\in\Pi(\mathcal A_G)
\end{align}
be an exact partition of $\Omega_G$ whose blocks belong to $\mathcal A_G$. Since only
\begin{align}
    A_{u,0},\quad A_{u,1},\quad A_{u,2}
\end{align}
contain the row $a_u$, the partition $\pi$ contains exactly one of these patterns. Define
\begin{align}
    \phi_\pi(u)=c
    \quad\Longleftrightarrow\quad
    A_{u,c}\in\pi.
    \label{partition coloring}
\end{align}
Suppose that $e=\{u,w\}\in E$ and
\begin{align}
    \phi_\pi(u)=\phi_\pi(w)=c.
\end{align}
Then both $A_{u,c}$ and $A_{w,c}$ contain $b_{e,c}$, contradicting the fact that the patterns in $\pi$ are pairwise disjoint. Hence, $\phi_\pi$ is a proper 3-coloring.

Conversely, let
\begin{align}
    \phi:V\to\mathcal C
\end{align}
be a proper 3-coloring. Select $A_{u,\phi(u)}$ for every $u\in V$. For every edge $e=\{u,w\}$, the colors $\phi(u)$ and $\phi(w)$ are distinct, so there is a unique color
\begin{align}
    c_e
    \in
    \mathcal C\setminus\{\phi(u),\phi(w)\}.
\end{align}
Select the filler $F_{e,c_e}$. These patterns cover every row in $\Omega_G$ exactly once and therefore form an exact partition. Consequently,
\begin{align}
    \Pi(\mathcal A_G)\neq\emptyset
    \quad\Longleftrightarrow\quad
    G\text{ is 3-colorable}.
    \label{partition coloring equivalence}
\end{align}

It remains to relate the existence of a coloring to perfect representability, which requires the full value vector to be generated by a convex combination of exact partitions.

Suppose that $\phi$ is a proper 3-coloring and define
\begin{align}
    \phi_t(u)
    :=
    \phi(u)+t
    \pmod 3,
    \qquad
    t\in\{0,1,2\}.
\end{align}
Let $\pi_t$ be the exact partition associated with $\phi_t$. Across the three partitions
\begin{align}
    \pi_0,\quad\pi_1,\quad\pi_2,
\end{align}
every vertex-choice pattern $A_{u,c}$ occurs exactly once, because each vertex receives each color exactly once. Moreover, for every edge, the missing color also cycles through $0,1,2$, so every filler pattern $F_{e,c}$ occurs exactly once.

Assign weight $1/3$ to each of the three partitions and weight zero to every other admissible partition. Then every pattern receives total weight $1/3$. Hence, for the reduced partition matrix $H$,
\begin{align}
    H\vec{x}
    =
    \frac{1}{3}
    \vec{1}_{|\mathcal A_G|}
    =
    \vec{v}.
\end{align}
By the characterization in Section~\ref{algorithmic results chapter}, $M(G)$ is perfectly representable.

Conversely, suppose that $M(G)$ is perfectly representable. Then there exists $\vec{x}\geq0$ such that
\begin{align}
    H\vec{x}
    =
    \frac{1}{3}
    \vec{1}_{|\mathcal A_G|}.
\end{align}
Since the right-hand side is nonzero, at least one coordinate of $\vec{x}$ is positive. Therefore, at least one admissible exact partition exists. By~\eqref{partition coloring equivalence}, $G$ is 3-colorable.

We have therefore shown that
\begin{align}
    G\text{ is 3-colorable}
    \quad\Longleftrightarrow\quad
    M(G)\text{ is perfectly representable}.
\end{align}
The construction has $|V|+3|E|$ rows and $3|V|+3|E|$ columns and can be carried out in polynomial time. Therefore, deciding\vadjust{\vskip 2.2pt} perfect representability is NP-hard.

We now prove membership in NP for an arbitrary rational singular conditional probability matrix encoded in binary. Merge columns with the same pattern as in Section~\ref{algorithmic results chapter}, and let $m$ be the number of distinct occurring patterns. By Theorem~\ref{LP algorithm theorem}, a yes instance admits a solution of $H\vec{x}=\vec{v}$ with $\vec{x}\geq0$. Choose one with the fewest positive weights. The corresponding columns of $H$ are linearly independent. Given a dependence, we could vary the weights along a linear dependence until one becomes zero, preserving both nonnegativity and $H\vec{x}=\vec{v}$. Thus at most $m$ weights are positive.

Let $r$ be the number of positive weights. They are determined by a nonsingular $r\times r$ binary submatrix and the corresponding $r$ entries of $\vec{v}$. Let $D$ be the product of the input denominators, so that $D\vec{v}$ is integral and $\log_2 D$ is at most the input bit length. By Cramer's rule, each weight can be written with numerator and denominator of absolute value at most $D r!$, since the submatrix is binary and $0\leq v_i\leq1$. Thus each weight has polynomial bit length. A certificate lists the $r$ partitions by their pattern indices, together with these weights. In polynomial time we can check that each listed partition covers every row exactly once, that all weights are nonnegative, and that they sum to $v_i$ over the partitions containing pattern $\vec{p}_i$, for every $i$. This proves NP membership. Together with the reduction, this shows that deciding perfect representability is NP-complete for rational singular conditional probability matrices encoded in binary, even when every row has exactly three nonzero entries, all equal to $1/3$.
\end{proof}

By Lemma~\ref{singularity lemma}, nonsingular conditional probability matrices are not perfectly representable. Since singularity can be checked in polynomial time, the same NP-completeness conclusion holds for arbitrary rational conditional probability matrices encoded in binary.

\subsection{Hardness Under Bounded Pattern Size}
\label{bounded pattern appendix subsection}

The reduction in Appendix~\ref{decision complexity proof subsection} works for arbitrary finite simple graphs. Restricting the source problem to graphs of maximum degree at most four gives a stronger hardness result. Graph 3-colorability remains NP-complete under this degree restriction~\cite{garey_johnson_stockmeyer_1976}.

For the vertex-choice patterns in~\eqref{vertex choice pattern},
\begin{align}
    |A_{u,c}|
    =
    1+\deg(u).
\end{align}
Therefore, when
\begin{align}
    \deg(u)\leq4,
\end{align}
we have
\begin{align}
    |A_{u,c}|\leq5.
\end{align}
Every filler pattern satisfies
\begin{align}
    |F_{e,c}|=1.
\end{align}

\begin{corollary}
\label{bounded pattern NP-hard corollary}
Deciding perfect representability is NP-complete for the class of singular conditional probability matrices satisfying all of the following properties:
\begin{enumerate}
    \item every positive matrix entry equals $1/3$;
    \item every row has exactly three positive entries;
    \item all occurring patterns have value $1/3$;
    \item every occurring pattern has size at most five.
\end{enumerate}
\end{corollary}

\begin{proof}
Apply the reduction in the proof of Theorem~\ref{perfect representability NP-hard theorem} to graphs of maximum degree at most four. The construction preserves the equivalence between 3-colorability and perfect representability. The bounds above show that every vertex-choice pattern has size at most five, while every filler pattern has size one. The four restrictions can be checked in polynomial time, and the same certificate proves membership in NP.
\end{proof}

%% file: appendixE.tex
\begin{proof}[Proof of Corollary~\ref{cor:graph-entropy}]
By the equivalent definition in~\cite{LiElGamal2018EGW}, the K{\"o}rner graph entropy of $G_{YX}$ under $p_Y$ is
\begin{align*}
    H_K(G_{YX},Y) = \min_{U:U-Y-X, H(Y|XU)=0} I(Y;U).
\end{align*}
For any $U$ in the minimization defining $H_K(G_{YX},Y)$, we have $U-Y-X$ and $H(Y|XU)=0$. Hence,
\[
I(Y;U)=I(X;U)+I(Y;U|X)=I(X;U)+H(Y|X).
\]
The minimum equals $H(Y|X)$ if and only if there exists such a $U$ with $I(X;U)=0$, that is, if and only if $p_{Y|X}$ is perfectly representable.
\end{proof}

\begin{proof}[Proof of Corollary~\ref{cor:privacy-funnel}]
For any $U$ with $X-Y-U$ and $I(X;U)=0$, we have $I(Y;U)=H(Y|X)-H(Y|XU)$. Thus, $I(Y;U)\ge H(Y|X)$ if and only if $H(Y|XU)=0$. Such a $U$ achieves zero leakage at disclosure level $H(Y|X)$ and is a perfect functional representation.

Moreover, $g_0(p_{XY})$, also denoted by $g_0(X;Y)$, is the maximum $t$ for perfect privacy~\cite{privacy_funnel2} and can be written as
\begin{align*}
    g_0(X;Y) = \max\{t \geq 0 : G_{\mathrm{PF}}\!\big(t,\, X \!\to\! Y\big)=0 \}.
\end{align*}
Since $I(Y;U)\le H(Y|X)$ under perfect privacy, $G_{\mathrm{PF}}\!\big(H(Y|X),\, X \!\to\! Y\big)=0$ if and only if $g_0(X;Y)=H(Y|X)$.
\end{proof}

\begin{proof}[Proof of Corollary~\ref{cor:excess-functional}]
For any $U$ with $I(X;U)=0$, the chain rule gives
\[
H(Y|U)-I(X;Y)=H(Y|XU)+I(X;U|Y).
\]
Thus, $\Psi(X \rightarrow Y)=0$ if and only if some such $U$ satisfies both $H(Y|XU)=0$ and $I(X;U|Y)=0$, precisely the conditions for $U$ to be a perfect functional representation.
\end{proof}

\begin{proof}[Proof of Corollary~\ref{cor:wyner}]
Following~\cite{yana}, we write
\begin{align*}
g_0(p_{XY}) &= \sup_{\substack{Z:I(X;Z)=0 \\ X-Y-Z}} I(Y;Z),\\
h_0(p_{XY}) &= \sup_{Z:I(X;Z)=0} I(Y;Z).
\end{align*}
Since $H(Y|X) = I(Y;Z) + I(X;Z|Y) + H(Y|XZ)$ when $I(X;Z) = 0$, both $g_0(p_{XY})$ and $h_0(p_{XY})$ are less than or equal to $H(Y|X)$ (equality holds if and only if $p_{Y|X}$ is perfectly representable).
In \cite{Zamani}, it is shown that $h_0(p_{XY}) = g_0(p_{XY})$ if $C(X,Y) = I(X;Y)$. Moreover, in \cite{yana}, it is shown that both $g_0(p_{XY})$ and $h_0(p_{XY})$ are equal to $H(Y|X)$\vadjust{\vskip 2.5pt} if $h_0(p_{XY}) = g_0(p_{XY})$. Combining these two results, we can say that $p_{Y|X}$ is perfectly representable if $C(X,Y) = I(X;Y)$.

As an example of a perfectly representable channel $p_{Y|X}$ for which $C(X,Y) \neq I(X;Y)$, we can consider a binary erasure channel with erasure probability $0<\epsilon<1/2$ and a uniform $p_X$. The erasure indicator is independent of $X$, is a function of $Y$, and together with $X$ determines $Y$. Thus, it is a perfect functional representation. For this channel, $C(X,Y) = 1$~\cite[Example~1]{kumar2014exact} and $I(X;Y) = 1-\epsilon$.
\end{proof}